\documentclass[a4paper,fleqn]{cas-dc}

\usepackage[numbers,sort&compress]{natbib}

\usepackage{array,graphicx}
\usepackage{lipsum}
\usepackage{etoolbox}
\usepackage{booktabs}
\usepackage{pdflscape}
\usepackage{longtable, tabularx}
\usepackage{verbatim}
\usepackage{gensymb}
\usepackage{tikz}
\usepackage{makecell} 
\usepackage{booktabs}
\usepackage{multirow}
\usepackage{multicol}
\usepackage{amsfonts}
\usepackage{amssymb,mathtools}
\usepackage{amsmath}
\usepackage{algpseudocode}
\usepackage[linesnumbered,ruled,vlined]{algorithm2e}
\usepackage[many]{tcolorbox}
\usepackage[thinc]{esdiff}
\usepackage{enumitem}
\usepackage[svgnames]{xcolor}
\usepackage{threeparttable}
\usepackage{wasysym}
\usepackage{hyperref}
\usepackage{textcomp}
\usepackage{caption}
\usepackage{subcaption}
\usepackage{float}
\usepackage{orcidlink}
\usepackage{url}
\usepackage{hyperref}

\def\tsc#1{\csdef{#1}{\textsc{\lowercase{#1}}\xspace}}
\tsc{WGM}
\tsc{QE}
\tsc{EP}
\tsc{PMS}
\tsc{BEC}
\tsc{DE}

\makeatletter
\def\NAT@def@citea{\def\@citea{\NAT@separator}}
\makeatother

\newcommand\clearrow{\global\let\rowmac\relax}
\clearrow

\ExplSyntaxOn
\NewDocumentCommand{\longdash}{ O{2} }
 {
  --\prg_replicate:nn { #1 - 1 } { \negthinspace -- }
 }
\ExplSyntaxOff

\newcommand{\mycomment}[1]{}

\ExplSyntaxOn

\NewDocumentCommand \vect { s o m }
 {
  \IfBooleanTF {#1}
   { \vectaux*{#3} }
   { \IfValueTF {#2} { \vectaux[#2]{#3} } { \vectaux{#3} } }
 }

\DeclarePairedDelimiterX \vectaux [1] {\lbrack} {\rbrack}
 { \, \dbacc_vect:n { #1 } \, }

\cs_new_protected:Npn \dbacc_vect:n #1
 {
  \seq_set_split:Nnn \l_tmpa_seq { , } { #1 }
  \seq_use:Nn \l_tmpa_seq { \enspace }
 }
\ExplSyntaxOff

\newcommand{\cs}[1]{\texttt{\symbol{`\\}#1}}

\makeatletter
\newcommand*{\compress}{\@minipagetrue}
\makeatother

\NewTColorBox{AlgBox}{ s O{!htbp} }{%
  floatplacement={#2},
IfBooleanTF={#1}{float*,width=\textwidth}{float},
  colback=gray!10, colframe=gray!10,
  sharp corners
}

\newcommand{\myfrac}[2]{%
    \setbox0\hbox{$#1$}        
    \dimen0=\wd0               
    \setbox1\hbox{$#2$}        
    \dimen1=\wd1               
    \ifdim\wd0<\wd1            
        \dfrac{#1\hfill}{#2}   
    \else                      
        \dfrac{#1}{#2\hfill}   
    \fi
}

\newcolumntype{?}{!{\vrule width 1.25pt}}

\newcolumntype{P}[1]{>{\centering\arraybackslash}p{#1}}

\begin{document}
\let\WriteBookmarks\relax
\def\floatpagepagefraction{1}
\def\textpagefraction{.001}

\shorttitle{}

\shortauthors{Michael Herman et~al.}

\title [mode = title]{Predictive calibration for digital sun sensors using sparse submanifold convolutional neural networks}                      

%
\author[1]{Michael Herman\,\orcidlink{0009-0001-1000-4553}}

\ead{hermanm@gatech.edu}



\affiliation[1]{organization={Aerospace Systems Design Laboratory, School of Aerospace Engineering, Georgia Institute of Technology},
    city={Atlanta},
    state={GA},
    postcode={30332}, 
    country={USA}}

\author[1]{Olivia J. Pinon Fischer\,\orcidlink{0000-0002-2233-1391}}

\ead{olivia.pinon@asdl.gatech.edu}

\author[1]{Dimitri N. Mavris\,\orcidlink{0000-0001-8783-4988}}

\ead{dimitri.mavris@aerospace.gatech.edu}


\begin{abstract}
Recent developments in AI techniques for space applications mirror the success achieved in terrestrial applications. Machine learning, which excels in data rich environments, is particularly well suited to space-based computer vision applications, such as space optical attitude sensing. Of these sensors, digital sun sensors (DSS) are one of the most common and important sensors for spacecraft attitude determination. The main challenge in using the DSS for attitude estimation are sensor errors, which limit the overall achievable estimation accuracy. However, the traditional sun sensor calibration process is costly, slow, labor-intensive and inefficient. These limitations motivate the use of AI techniques to enable more accurate and efficient DSS calibration. 

The objective of this work is to develop an end-to-end predictive calibration methodology for digital sun sensors to solve 2-axis state estimates utilizing a sparse submanifold convolutional neural network (SSCNN). We find that the proposed framework can achieve state-of-the-art performance on synthetic data with a mean accuracy of 0.005° for the two sun angle estimates. Furthermore, the model is highly capable of implicitly learning complex noise patterns and handling mixed noise types, thereby greatly improving the model robustness and accuracy to real-world applications. The main contributions of this work are: (1) the first application (to our knowledge) of a CNN regression model to the problem of DSS predictive calibration, (2) the introduction of a fused end-to-end training approach for DSS calibration, (3) the creation of a publicly available physics-informed synthetic dataset and simulation for DSS training images, and (4) the evaluation of the performance of the deep learning approach for various mask configurations.
\end{abstract}



\maketitle

\section{Introduction}\label{sec:intro}

Attitude sensors determine the spacecraft orientation through the sensing of an astronomical object. The Sun and fixed stars are the two primary astronomical sensing objects. Attitude sensors are critical components for the survival and state knowledge of spacecraft. The sun sensor, magnetometer, star sensor, and Earth sensor make up this category. Of these, sun sensors are one of the most common and important sensors for small satellite attitude determination \cite{You2011}.

Nearly all low-Earth orbiting small satellites employ sun sensors as part of the attitude sensor package, which determine the satellite attitude by measuring the Sun vector relative to the satellite coordinates \cite{KeQiang2020}. The scope of this research extends beyond space, such as terrestrial navigation and ground applications on other planets. For example, sun sensors can enable more accurate rover heading for planetary exploration \cite{Furgale2010}, terrestrial navigation can be improved through geolocation \cite{Barnes2014} and by correcting for cloudy environments \cite{Wang2024}, and 3D target localization \cite{Zhang2022a} can be achieved with multi-aperture sensors. The two main categories of sun sensors are analog and digital, however digital sun sensors (DSS) constitute the largest fraction of modern high-accuracy sun sensors.

The main challenge for using DSS for attitude estimation is sensor errors. These errors limit the overall achievable attitude estimation accuracy. During development and in-orbit operation, the sensors are affected by numerous sources of uncertainties, including manufacturing, environmental, interference sources, or those inherent to the sensor architecture. The sun sensor calibration process is particularly difficult due to the complex nature of the uncertainties involved. In the survey by \citet{Herman2025c}, the authors identified key challenges for sun sensor calibration algorithms relevant to this study.

\textbf{The authors identified three key challenges:} \cite{Herman2025c}
\begin{enumerate}
    \item There is a lack of publicly available datasets to test and train sun sensor calibration algorithms.
    \item Models are often tightly coupled to specific architectures and require manual error characterization.
    \item Feature extraction improvements face diminishing returns and are limited in their ability to leverage rich feature spaces for calibration mapping.
\end{enumerate}

This work aims to address each of these three key challenges as a primary contribution. As such, these challenges motivate the development of enhanced calibration techniques to improve upon the current state of the art of sun sensor calibration and minimize uncertainty over the sensor lifecycle. However, traditional sun sensor calibration techniques suffer from laborious experiments to gather data, model uncertainties, and feature extraction limitations. Those challenges are discussed in more detail below. 

\textbf{Process limits.} The sun sensor calibration process is mired by labor-intensive and time-consuming experiments to gather training data. As a result, there is a lack of large and comprehensive datasets of real sensor images or sufficiently realistic synthetic data.

\begin{figure*}[ht]
\noindent\rule{\textwidth}{1.25pt} 
\centerline{\includegraphics[width=\textwidth]{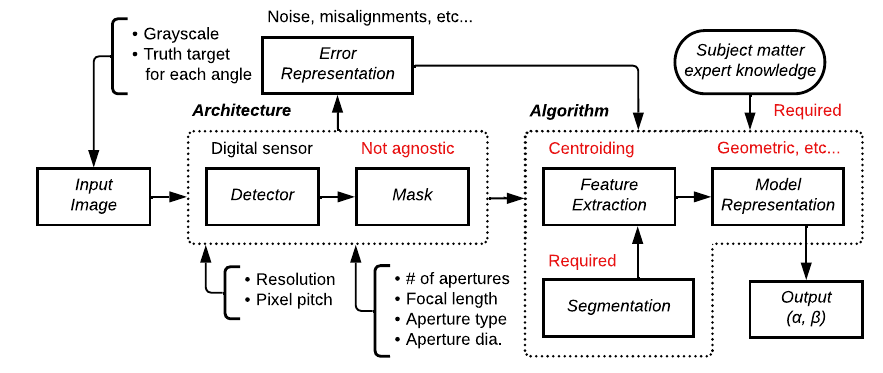}}
\noindent\rule{\textwidth}{1.25pt} 
\caption{Traditional digital sun sensor calibration process flow. We note the limitations that our proposed method overcomes in red text.}
\label{fig:tradcalflow}
\end{figure*}

\textbf{Model limits.} Currently, there is no one size fits all solution to DSS calibration. Traditional calibration is inflexible and tied to specific sensor architectures. This lack of model generalizability requires the development of new models for each unique mask configuration. Furthermore, manual formulation of the sensor parameters and error sources is required to develop the models. Traditional calibration requires separate algorithms for feature capture, segmentation, and correlation mapping model representations, thereby increasing the algorithm computational overhead and implementation complexity. The challenges of the traditional DSS calibration process are outlined in Figure \ref{fig:tradcalflow}.

\textbf{Feature extraction limits.} Conventional feature extraction is typically simplified by reducing the problem to single or multi-point detection. As a result, the richness of the full feature space is lost. For example, DSS compute sparse features like centroids rather than capturing the full feature space. In addition, sensor noise floor challenges limit the sub-pixel accuracy of centroiding methods. Image anomalies such as debris and mask pollution degrade sensor performance and require additional detection schemes to account for. Diffraction effects are often mitigated in design rather than accounted for as a learned feature during calibration.

\textbf{Proposed methodology.} This work proposes to address the aforementioned challenges by leveraging convolutional neural networks (CNNs) trained with synthetic images. CNNs have demonstrated significant value in the space domain, enabling advancements in star trackers \cite{Zhao2024}, rendezvous operations \cite{Sharma2020}, hazard avoidance \cite{Tomita2022}, and pose estimation \cite{Pasqualetto2021}. Beyond space applications, CNNs have been instrumental in fields such as medical diagnostics \cite{Fiorentino2021}, optical measurement \cite{Liu2021}, astronomy \cite{Zhu2020}, and orientation estimation \cite{Tsai2020}. While recent research has explored the use of deep neural networks (DNNs) for calibrating analog sun sensors \cite{Soken2023,Sun2023}, the application of CNNs to digital sensors remains largely unexplored. To address this gap, this study introduces an end-to-end training approach for a unified model-feature calibration CNN. This model is designed to simultaneously perform feature extraction and correlation mapping, enabling predictive calibration for DSS.

\section{Related work}

In this section, we outline the surrounding sun sensor calibration literature. We begin with a brief review of classic sun sensor calibration techniques, followed by an examination of the state of the art in AI-based sun sensor calibration methods, and then close with a discussion on the various applications of simulated data for sun sensor calibration.

\subsection{Classical sun sensor calibration}
Over the years, researchers have developed a wide range of digital sun sensor calibration techniques. Traditionally, the calibration process has centered around two core algorithmic components: model representation and feature extraction. \citet{Herman2025c} conducted a systematic mapping and survey of calibration techniques, covering approaches such as lookup tables (LUTs), non-physical models, geometric methods, physics-informed models, multiplexing strategies, and neural networks. This section presents a brief summary of the calibration techniques identified in the systematic mapping conducted in the aforementioned survey.

LUTs are simple models that store the sensor output response or the associated fitted coefficients in a lookup table for later interpolation. Non-physical models involve the fitting of algebraic or transcendental regression models to measured sensor observations. These methods lack physical interpretability of the system from the model parameters and include the linear fit, polynomial, trigonometric, Fourier series, and sigmoid model representations. Geometric model representations involve parameters that are physically interpretable and account for the physical geometry of the sensor in the inverse function. These models are typically tied to a specific sensor architecture configuration. Some geometric models include the standard projection model, LSQ model, architecture specific models (QPD, slit, multi-slit, V-slit, N-slit, camera, etc), and multi-sensor fusion models (basic, solar panel, pyramidal, panoramic, etc).

Physics-informed models involve the compensation of errors derived from physical processes inherent to the sensor configuration. Some forms of these physical errors include environmental, manufacturing and assembly, optical, electrical, and interference effects. Multiplexing model representations use coding pattern rules related to the mask configuration to enable sun vector estimation with very high accuracy and a large field of view. The two main multiplexing model types are periodic and coded variants.

\subsection{AI-based sun sensor calibration}

Next, we review the general use of machine learning (ML) for space attitude sensor algorithms, followed by the early applications of ML for sun sensors, and close with more modern implementations of DNNs for sun sensor calibration.

\textbf{The case for AI in space attitude sensors.} In the recent years, machine learning approaches such as artificial neural networks (ANN), deep neural networks (DNN), and convolutional neural networks (CNN) have become essential tools for the aerospace research community. \citet{Kothari2020} argues that deep learning is being increasingly applied for space applications and on-device deep learning can improve the operation of space sensors. In particular, star trackers, Earth cameras, and sun sensors have all demonstrated improvements from the application of machine learning methods. \citet{Zapevalin2022} computed the star tracker centroid with a CNN. In addition, \citet{Zhao2024} applied a CNN to star detection and centroiding. \citet{Koizumi2018,Iwasaki2019,Kikuya2023} proposed the DLAS, which is a Earth camera sensor that estimates the attitude by classifying the scene with a neural network and applying template matching.

\textbf{Early AI use for sun sensor calibration.} Neural network models are a machine learning approach that use interconnected artificial neurons arranged in a layered structure to approximate the structure and function of the human brain. This representation is powerful because it can capture relationships between large volumes of sensor data. Early machine learning approaches for sun sensors were limited to shallow networks and sparse feature sets such as centroids. \citet{Rufino2012} implemented a multilayer feed-forward neural network to estimate the sun vector for a multi-aperture digital sun sensor by training on the average centroid coordinates.

\textbf{DL methods for sun sensors.} Despite the growing application of machine learning techniques to space sensors, limited coverage has been provided to sun sensor calibration. Recently, deep learning has been investigated to calibrate sun sensors, however these methods have primarily focused on analog sensors \cite{Sozen2021,Soken2023,Sun2023}. \citet{Sozen2021,Soken2023} applied a DNN to analog sun sensors for albedo correction. In addition, \citet{Sun2023} calibrated an analog sun sensor with cubic surface fitting and a DNN.

\subsection{Leveraging synthetic data for sun sensors} When limited real data is available, such as when collecting space-based imagery, synthetic data has been successfully leveraged during pre-training to improve model performance on the associated real-world dataset \cite{Hansen2017,Isele2017,Wu2018,Allworth2021}. Transfer learning is applied to fine-tine the pre-trained model on a smaller real set of images. This method is most effective when there is limited availability of the real-world target set and the target set is similar to the pre-training data \cite{Allworth2021}. We have identified five different cases of simulation use in sun sensor calibration literature, which include: (1) synthetic illumination images for centroiding, (2) simulated sensor response, (3) synthetic data for neural network-based sun sensors, (4) simulated uncertainties, and (5) simulated physics. In this section we outline each identified simulation variant with an example case-study from the literature.

\textbf{Synthetic illumination images for centroiding.} The first identified use of simulations for sun sensors is the generation of synthetic illumination images for model training, in which we found 12 examples in the literature \cite{Merchan2023,Bao2022,Enright2006b,Enright2007a,Enright2008a,Bolshakov2020,He2005,Buonocore2005,Antonello2018,Enright2007b,Enright2008b,Enright2008c}. This approach usually involves the generation of illumination spots from diffraction simulations or optical approximations. An example of this case is in the work by \citet{Enright2008a}, in which a simulation was used to model Fresnel diffraction, pixel quantization, solar divergence, and detector noise in the synthetic images.

\textbf{Simulated sensor response.} The second and most common identified use of simulations for sun sensors is to model the mapping of the sensor response, in which we found 13 examples in the literature \cite{Salazar2024,Wei2011,Mafi2023,KeQiang2020,Yousefian2017,Fan2015,Delgado2010,Delgado2012a,Delgado2012b,Strietzel2002,Farian2015,Diriker2021,Fan2013}. Usually this is done through the formulation of the sensor behaviors such as architecture geometry, offset parameterization, or detector modeling. An example of this approach is in the work by \citet{KeQiang2020}, in which an error compensation model is developed for a slit type digital sun sensor.

\textbf{Synthetic data for NN-based sun sensors.} The third identified use of simulations for sun sensors is the training of NN-based sensors with synthetic data. Since neural networks (NN) require large amounts of data for training, it is often necessary to utilize synthetic datasets in order to accommodate this requirement. While the use of NNs for sun sensor calibration is still nascent, there are 2 examples in the literature that rely on simulations for data generation \cite{Soken2023,Sozen2021}. For example, in the work by \citet{Sozen2021}, an ANN is trained for the measurement correction of analog sun sensors under albedo errors.

\textbf{Simulated uncertainties.} The fourth identified use of simulations for sun sensors is the generation of synthetic uncertainties. Errors can occur in flight due to uncertainties from the detector or mask. These errors are often hard to replicate without simulated data due to their anomalous nature and potential for sensor damage. Some examples include the deterioration of the sensor mask due to space debris, mask pollution, and detector aging. For this case, we refer to the study by \citet{You2011}, in which synthetic deteriorated illumination images are used to test the robustness of the algorithm when some apertures are missing and the image is polluted.

\textbf{Simulated physics.} The fifth and last simulation variant we have identified in the literature is the use of physics simulations, where we have found 4 examples in the literature \cite{Zhang2022a,Guler2021,Fan2016,Wei2014}. These simulation are used to either validate the experimental findings or as a synthetic substitute when experimental results are challenging to obtain. Some examples of synthetic physics-informed data include: diffraction patterns, albedo effects, and reflectance. For example, in the work by \citet{Zhang2022a} the authors simulate the spot profile based on the Fresnel-Kirchhoff diffraction formula.

\section{Contributions and paper outline}
With these considerations in mind, we explore the application of deep learning to DSS calibration. To the best of our knowledge, this is the first study to investigate the use of regression-based CNNs for predictive calibration of DSS, and in particular, to explore the application of sparse submanifold convolutional neural networks (SSCNNs) within this context. Our approach was developed and evaluated using a synthetic dataset generated through physics-informed simulations.

\textbf{The primary contributions of this work are:}
\begin{enumerate}
\item To the best of our knowledge, this research is the first application of a CNN regression model to address the problem of predictive calibration for DSS.
\item We propose an end-to-end training approach for a single CNN, designed to concurrently handle feature extraction and correlation mapping for DSS.
\item We generate a synthetic dataset of physics-informed digital sun sensor images, augmented with real sensor noise and thresholding, for training purposes. This approach greatly reduces the labor of experiments and manual labeling efforts.
\item We evaluate the performance of the deep learning approach under different mask configurations.
\end{enumerate}

This article is structured as follows: Section \ref{sec:measmodel} presents a formulation of the concept of DSS operation; Section \ref{sec:dataset} presents the physics-informed sun sensor simulation and outlines the data generation process for the model training and validation; Section \ref{sec:methods} presents the workflow of the proposed sun sensor calibration method, introduces the SSCNN algorithm, details the proposed SSCNN regression implementation, and outlines the model training and validation results; Section \ref{sec:result} presents the testing and robustness results of the trained models; Section \ref{sec:discuss} discusses key findings, highlights potential limitations, and recommends future research directions; finally, Section \ref{sec:conc} concludes the paper.

\section{Measurement model for sun sensors}\label{sec:measmodel}

\begin{figure}[ht]
	\centering
		\includegraphics[width=0.5\textwidth]{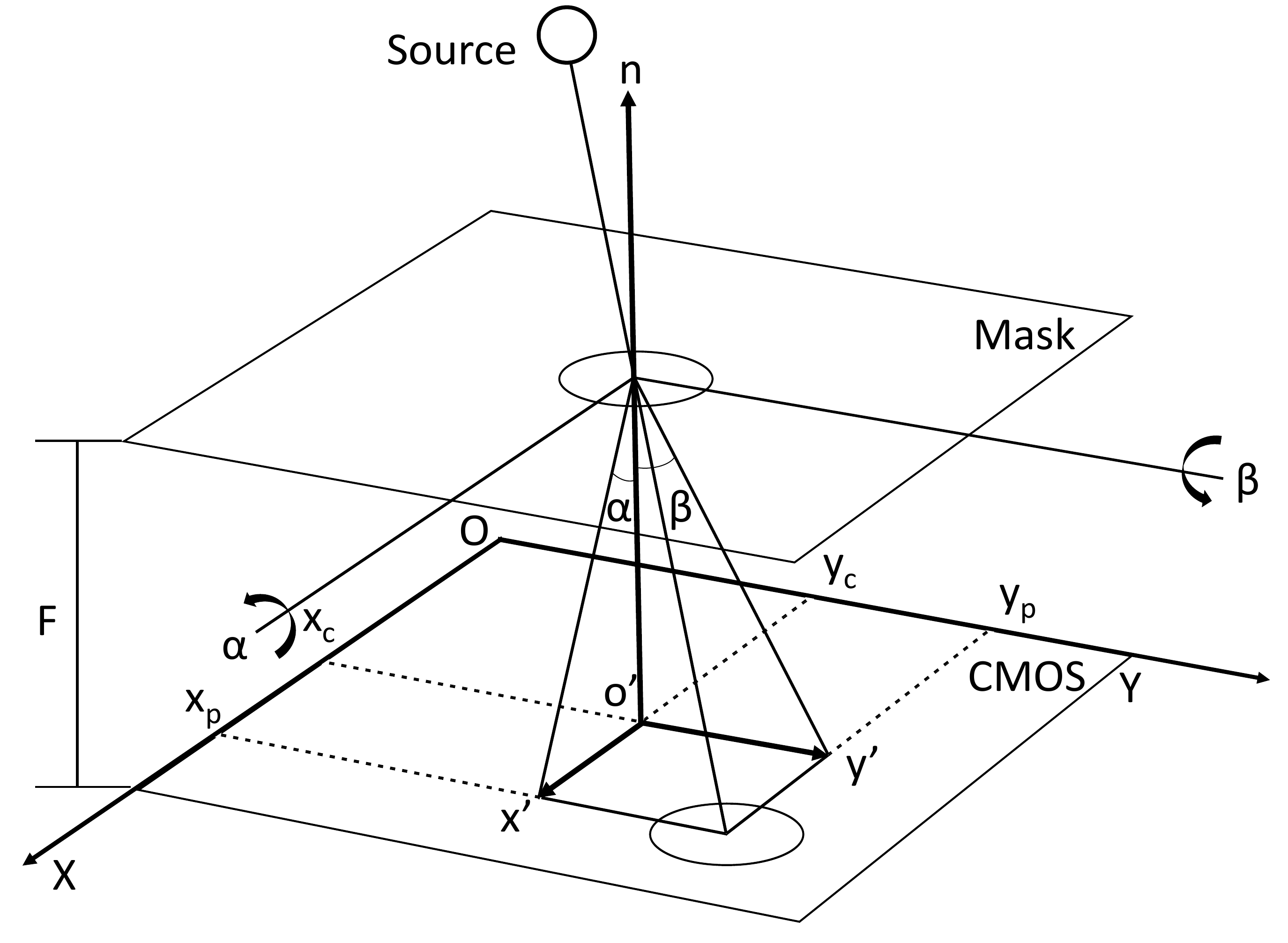}
	\caption{Ideal single-aperture sensor operating principle. The two sun angles $\alpha$ and $\beta$ are the values estimated in this study.}
	\label{fig:operbw}
\end{figure}

This section provides an overview of the sun sensor concept, its operating principles, and its architectural components. The sun sensor is employed to detect the satellite’s attitude angle with respect to the sun. Its working principle is illustrated in Figure \ref{fig:operbw}. With the angle information, the satellite’s location in space can be achieved by a dedicated algorithm. 

The working principle of the sun sensor is as follows. The sun sensor, which is placed on the satellite, has a thin mask above the chip surface. This mask has a pinhole on the top. In this way, the sunlight goes through this pinhole leaving a sun spot on the image sensor array. As the sun light passes through the hole, a nearly circular spot is formed on the sensing surface of the detector.

Due to the distance F between the focal plane and the mask, representing the focal length of the system, the spot position changes as the illumination direction varies, thus allowing its measurement on the basis of spot center position. The image sensor within the sensor package reads out this sun spot’s location on the focal plane. With this location information, the processing algorithm then calculates the attitude angle of the satellite with respect to the sun.

\section{Methodology}\label{sec:dataset}

In this section, we detail the experimental setup, including the creation of a synthetic dataset, discuss the data acquisition process, and review the data augmentation types.

\subsection{Synthetic dataset creation}

Convolutional neural networks (CNNs) require substantial labeled training data and computational effort to achieve optimal performance, primarily due to their large number of parameters \cite{Allworth2021}. For instance, the ResNet-34 model \cite{He2015} contains 63.5 million parameters and is trained on the ImageNet dataset \cite{Deng2009}, which includes 1.2 million images. Generally, increasing the number of training images improves model performance. However, gathering large datasets of labeled images is often impractical due to the significant time and cost involved. This challenge is particularly evident in the collection of calibration images during ground testing, which requires labor-intensive and expensive experiments using solar simulators. Furthermore, in-flight calibration is even more difficult due to the low availability of space imagery \cite{Sharma2018}, in which the acquisition and labeling of sun sensor images is required during space operations.

As a result, synthetic data was chosen as a more practical alternative. However, no publicly available dataset of digital sun sensor illumination images suitable for benchmarking calibration algorithms was found. To address this gap, a physics-informed simulation was developed to generate realistic synthetic data and streamline the data acquisition process. The development of a publicly available dataset and an open-source simulation framework for generating digital sun sensor images represent the third contribution of this study. 
The process by which the dataset was generated is discussed below. More detail can be found in the work by \citet{Herman2025a}.


We developed a physics-informed numerical simulation framework using Ansys Zemax OpticStudio and the MATLAB programming environment. OpticStudio was chosen as the simulator because of its strengths in sequential design, capability to create off-axis systems, and seamless integration with MATLAB via the ZOS-API. Furthermore, MATLAB was selected as the programming environment due to its robust data visualization capabilities and the extensive resources available for working with the ZOS-API. A large number of diffraction images were generated for a given mask configuration to serve as training data. The dataset was then labeled with the corresponding truth angles and enhanced using data augmentation techniques.

\textbf{Sensor parameters.} Before outlining the OpticStudio development process, the sensor optical design process will first be discussed. Here, two design configurations are proposed for analysis in the study. The proposed designs include: single aperture and multi-aperture mask patterns. Each aperture is a pinhole shape with a defined thickness. The multi-aperture design is comprised of a $3\times3$ array of 9 apertures with a given spacing. The related optical parameters for each mask design are presented in \ref{tab:opticalparam}.

We selected the Sony IMX477 sensor as the notional detector for simulation purposes. This detector was selected due to its low cost, availability, and ease of implementation on a CubeSat platform. Since the sensor is not monochromatic, we modeled the Bayer filter to account for the change in spectral response. We also used the sensor noise profile to augment the illumination images with real sensor noise. The IMX477 detector parameters are presented in Table \ref{tab:detectorparam}.

\begin{figure*}
\centering
\begin{subfigure}{.49\textwidth}
    \centering
    \includegraphics[width=\linewidth]{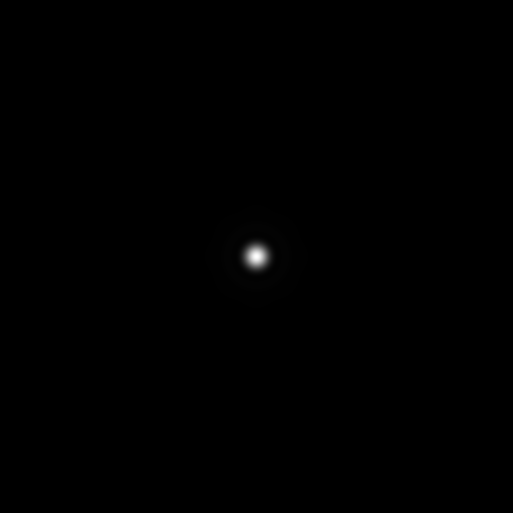}  
    \caption{Single-aperture mask illumination image at ($\alpha,\beta$) = 0}
    \label{fig:spota}
\end{subfigure}
\hfill
\begin{subfigure}{.49\textwidth}
    \centering
    \includegraphics[width=\linewidth]{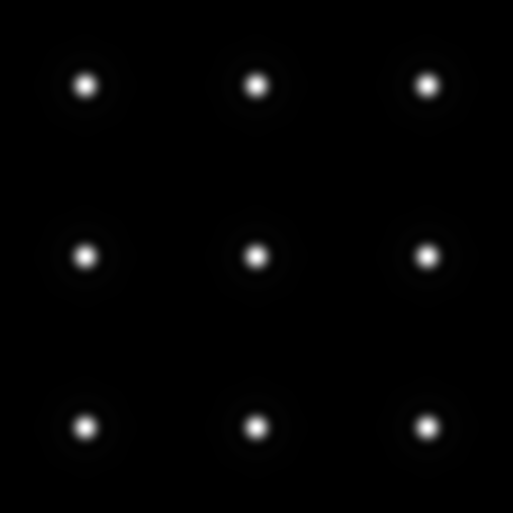}  
    \caption{Multi-aperture mask illumination image at ($\alpha,\beta$) = 0}
    \label{fig:spotb}
\end{subfigure}
\caption{Single and multi-aperture mask configuration illumination patterns at ($\alpha,\beta$) = 0. An example of the training images used for this study.}
\label{fig:spotab}
\end{figure*}

\begin{table}[htb]
\caption{Proposed system optical parameters.}
\label{tab:opticalparam}
\centering
\setlist[itemize]{wide, leftmargin=*, noitemsep, after=\vspace*{-\topsep}}
\setlength{\extrarowheight}{3pt}
\setlength{\tabcolsep}{3pt}
\newlength\ct
\setlength\ct{\dimexpr .1666\textwidth -2\tabcolsep}
\noindent\begin{tabular}{>{\compress\rowmac}p{1.33\ct}|>{\compress\rowmac}P{0.83\ct}|>{\compress\rowmac}P{0.83\ct}<{\clearrow}}
\Xhline{1.25pt}
\textbf{Parameter} & \textbf{Single-Aperture} & \textbf{Multi-Aperture}\\
\hline
No. of apertures & 1 & 9\\
Diameter ($\mu$m) & 100 & 100\\
Focal length (mm) & 5 & 5\\
Thickness (mm) & 0.03 & 0.03\\
\Xhline{1.25pt}
\end{tabular}
\end{table}

\begin{table}[htb]
\caption{Proposed system detector parameters.}
\label{tab:detectorparam}
\centering
\setlist[itemize]{wide, leftmargin=*, noitemsep, after=\vspace*{-\topsep}}
\setlength{\extrarowheight}{3pt}
\setlength{\tabcolsep}{3pt}
\setlength\ct{\dimexpr .25\textwidth -2\tabcolsep}
\noindent\begin{tabular}{>{\compress\rowmac}p{\ct}|>{\compress\rowmac}p{\ct}<{\clearrow}}
\Xhline{1.25pt}
\textbf{Parameter} & \textbf{Value}\\
\hline
Bit depth (bits) & 8\\
Pixel pitch ($\mu$m) & 1.55\\
Resolution & $1\times512\times512$\\
Angular resolution (mrad) & 0.31\\
Read Noise (e-) & 3\\
Full Well Capacity (e-) & 8180\\
\Xhline{1.25pt}
\end{tabular}
\end{table}

\textbf{Zemax configuration.} Next, the OpticStudio development process is discussed, in which the system configuration, optical surface ordering, analysis, and programming link are outlined. We developed the simulation and modify system parameters fully in the MATLAB environment using the ZOS-API interface. An in-depth overview of the programming workflow is discussed at the end of the section.

Sequential ray-tracing mode was selected for the optical system configuration, as it is computationally fast and well suited to analyzing imaging and afocal systems. In sequential mode, the incident rays are traced through a sequence of surfaces, while traveling from the object to image plane. In this mode, each ray must impinge a surface once in the order defined by the sequence of surfaces through the Lens Data Editor (LDE) interface. One downside of sequential mode is that it has less support for complex sources. To alleviate this limitation, the use of mixed mode can be enabled in Zemax in order to utilize both sequential surfaces and non-sequential objects in the same scenario. Therefore, complex sources such as the Sun as an extended source can be implemented.

\textbf{System Explorer settings.} The primary settings of interest in the System Explorer include aperture, fields, and wavelength. In the System Explorer, the aperture type is an entrance pupil diameter with an aperture value of 0.1mm for the single and multi-aperture configurations. Afocal image space mode is applied in the simulation since the optical system will not form an image on the image side. Furthermore, when afocal mode is selected, the evaluation of the image side is calculated in angular units as microradians. The extended nature of the Sun can be approximated using "Source Two Angle" in mixed mode, in which the source is modeled as a point source with a defined angular spread by setting the half-angle to 0.27°.

An accurate spectral representation is important since the wavelength affects the pinhole resolution, diffraction point spread function (PSF), and resulting Airy disk size. Here, we explain how both the solar spectral irradiance and sensor spectral sensitivity are accounted for to create a physically accurate adjusted spectrum for OpticStudio wavelengths. The adjusted spectrum is weighted by source curves, detector curves, and transmission, which are multiplied all together.

The source spectrum was customized as an approximate AM0 spectrum to simulate the space solar characteristics. The AM0 2000 ASTM Standard Extraterrestrial Spectrum Reference E-490-00 is the standard spectrum for space applications. The spectrum data used for this study was retrieved from a revised extraterrestrial solar spectrum proposed by \citet{Gueymard2018}.

The detector spectral sensitivity of the Sony IMX477 sensor was obtained and traced from datasheets. The nominal sensor spectral response was approximated by accounting for the Bayer color filter array (CFA) response spectra as a 1:2:1 RGB weighting, respectively. Therefore, the CFA responses are added as $R + 2G + B$ to approximate the nominal system sensitivity and account for the number of pixels of each color filter.

\begin{figure}
	\centering
        \includegraphics[width=0.5\textwidth]{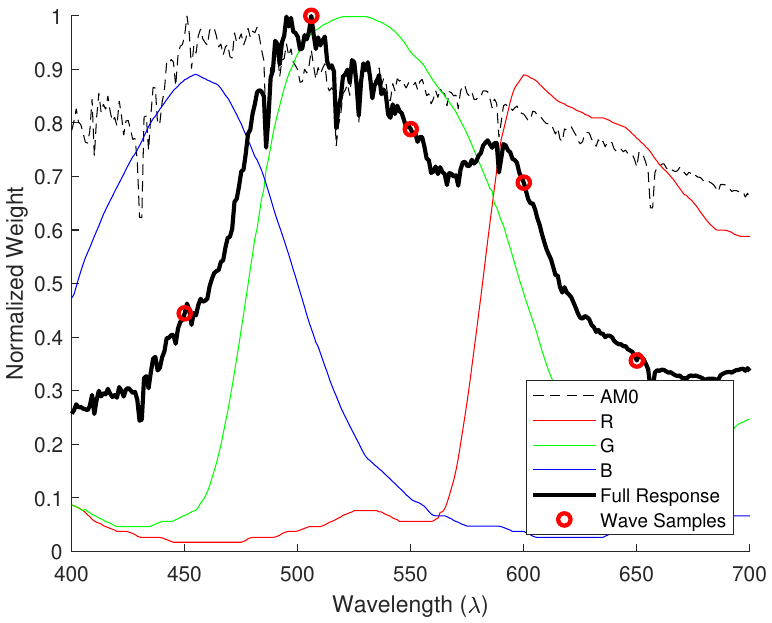}
	\caption{Adjusted spectral response.}
	\label{fig:adjspectrum}
\end{figure}

The adjusted spectral response is then calculated by multiplying the source curve by the nominal detector curve. The spectrum was limited to the visible wavelength range of 400nm-700nm. The adjusted spectrum is approximated by five spectral weightings along the distribution for use in OpticStudio and created in MATLAB using ZOS-API. The adjusted spectral response diagram is presented in Figure \ref{fig:adjspectrum}, where the sensor RGB response and AM0 spectrum are overlayed. In addition, the five sampled spectral weights used in the System Explorer wavelength settings are listed in Table \ref{tab:spectralparam}.

\begin{table}[htb]
\caption{OpticStudio custom AM0 spectral parameters.}
\label{tab:spectralparam}
\centering
\setlist[itemize]{wide, leftmargin=*, noitemsep, after=\vspace*{-\topsep}}
\setlength{\extrarowheight}{3pt}
\setlength{\tabcolsep}{3pt}
\setlength\ct{\dimexpr .167\textwidth -3\tabcolsep}
\noindent\begin{tabular}{>{\compress\rowmac}P{0.5\ct}|>{\compress\rowmac}P{1.25\ct}|>{\compress\rowmac}P{1.25\ct}<{\clearrow}}
\Xhline{1.25pt}
\textbf{No.} & \textbf{Wavelength $\lambda$ (nm)} & \textbf{Normalized Weight}\\
\hline
1 & 450.0000 & 0.4444\\
2 & 506.0000 & 1.0000\\
3 & 550.0000 & 0.7884\\
4 & 600.0000 & 0.6884\\
5 & 650.0000 & 0.3562\\
\Xhline{1.25pt}
\end{tabular}
\end{table}

\textbf{Lens Data Editor (LDE) configuration.} Here, we outline the OpticStudio surface components as entered in the Lens Data Editor (LDE) interface and list them in Table \ref{tab:lde}. The first surface, \texttt{Surface\#0}, is a source object located at infinity, which approximates the Sun as a point source. The second surface, \texttt{Surface\#1}, is a coordinate break to allow for tilting of the optical axis through the two-axis sun angles. Since this is an off-axis system, it is important to be able to model the effects of tilted rays on the detector plane. The third surface, \texttt{Surface\#2}, is the upper aperture of the pinhole mask and functions as the primary stop surface of the optical configuration. The fourth surface, \texttt{Surface\#3}, is the lower aperture of the pinhole mask and is placed to simulate a mask with a thickness offset from the upper surface. Finally, the fifth and last surface, \texttt{Surface\#4}, is an image plane for capturing the incident light on the detector.

\begin{table*}[htb]
\caption{OpticStudio Lens Data Editor (LDE).}
\label{tab:lde}
\centering
\setlist[itemize]{wide, leftmargin=*, noitemsep, after=\vspace*{-\topsep}}
\setlength{\extrarowheight}{3pt}
\setlength{\tabcolsep}{3pt}
\setlength\ct{\dimexpr .125\textwidth -2\tabcolsep}
\noindent\begin{tabular}{>{\compress\rowmac}P{0.5\ct}|>{\compress\rowmac}P{1.5\ct}>{\compress\rowmac}P{\ct}|>{\compress\rowmac}P{\ct}|>{\compress\rowmac}P{\ct}|>{\compress\rowmac}P{\ct}|>{\compress\rowmac}P{\ct}|>{\compress\rowmac}P{\ct}<{\clearrow}}
\Xhline{1.25pt}
\textbf{\#} & \multicolumn{2}{c|}{\textbf{Surface Type}} & \textbf{Radius} & \textbf{Thickness} & \textbf{Mech Semi-Dia} & \textbf{Tilt About X} & \textbf{Tilt About Y}\\
\hline
0 & OBJECT & Standard & Infinity & Infinity & 0.000 & -- & --\\
1 & \multicolumn{2}{c|}{Coordinate Break} & -- & 0.000 & -- & $\alpha$ & $\beta$\\
2 & STOP (aper) & Standard & Infinity & 0.030 & 0.050 & -- & --\\
3 & (aper) & Standard & Infinity & 4.985 & 0.050 & -- & --\\
4 & IMAGE & Standard & Infinity & -- & 5.5 & -- & --\\
\Xhline{1.25pt}
\end{tabular}
\end{table*}

The Huygens PSF was used to calculate the diffraction pattern since the Huygens PSF calculation automatically accounts for complex image surface tilt, while the FFT PSF calculation does not. The image surface is significantly tilted with respect to the chief ray for large boresight angles. The analysis parameters for the Huygens PSF generation are listed in Table \ref{tab:analysisparam}.

\begin{table}[htb]
\caption{OpticStudio Huygens PSF analysis parameters.}
\label{tab:analysisparam}
\centering
\setlist[itemize]{wide, leftmargin=*, noitemsep, after=\vspace*{-\topsep}}
\setlength{\extrarowheight}{3pt}
\setlength{\tabcolsep}{3pt}
\setlength\ct{\dimexpr .125\textwidth -2\tabcolsep}
\noindent\begin{tabular}{>{\compress\rowmac}p{1.5\ct}|>{\compress\rowmac}P{0.83\ct}|>{\compress\rowmac}P{0.83\ct}|>{\compress\rowmac}P{0.83\ct}<{\clearrow}}
\Xhline{1.25pt}
\textbf{Parameter} & \textbf{Single-Aperture} & \textbf{Multi-Aperture} & \textbf{Coded-Aperture}\\
\hline
Pupil Sampling & $32\times32$ & $32\times32$ & $32\times32$\\
Image Sampling & $128\times128$ & $128\times128$ & $64\times64$\\
Image Delta & 5 & 5 & 5\\
Wavelength & All & All & All\\
Show As & Inverse Greyscale & Inverse Greyscale & Inverse Greyscale\\
\Xhline{1.25pt}
\end{tabular}
\end{table}

The Zemax OpticStudio ZOS-API was used to interface with the MATLAB programming environment for synthetic training data generation. The physics-informed simulation code is available on \href{https://doi.org/10.5281/zenodo.16579563}{Zenodo} \cite{Herman2025g}. In addition, the full list of associated software and data repositories is detailed in Appendix \ref{sec:appendixa}.

\begin{figure}
	\centering
        \includegraphics[width=0.5\textwidth]{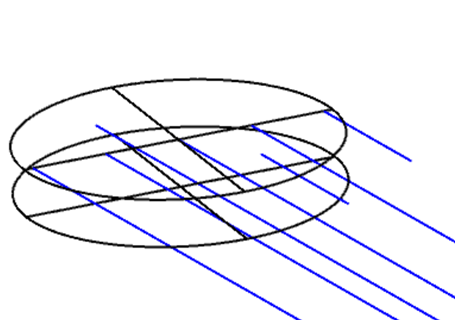}
	\caption{OpticStudio simulation.}
	\label{fig:zemax}
\end{figure}

\subsection{Data acquisition} \label{sec:daq}

The first step in the training process is data acquisition, where the synthetic data is generated from physics-informed simulations and assigned targets and sub-FOV classes. Our data set contains two parts, one is the original images, and the other are the two-axis truth targets stored in a CSV file. The generated images are $1\times512\times512$ to speed up training on the 4090 GPU. We generate images for two mask configurations including single-aperture and multi-aperture designs.

We divide the data into a training set, a validation set, and a testing set. The training set is used for supervised learning to train the model weights. The validation set is used to verify the model generalization and tune hyperparameters. Finally, the testing set is used to assess the model performance on augmented and non-augmented data for the two mask configurations.

The data is split into the three sets at a ratio of 8:1:1, respectively. Moreover, we assign data to the three sets at random, while ensuring that the aforementioned ratio is maintained. The total dataset size is 789,507 for a single mask configuration with augmented data. Furthermore, since there are three augmentation image types (original, noise, thresholding), the size of the non-augmented dataset is 263,169. However, training revealed that approximately 65,536 images were sufficient for achieving good convergence. This corresponds to a proposed sampling density of about $65,536/(512\times512)=1/4$, significantly reducing the amount of data required from simulation or ground testing to apply this method.

\subsection{Data augmentation} \label{sec:daug}

The data augmentation process introduces feature variance into the training sample population. Each pre-processed image is processed by an algorithm that applies a filter to introduce small perturbations in the image space. Examples of such perturbations include noise injection and thresholding. It is important that these modifications remain naturally invariant to the underlying physics of the sensor.
Therefore, some understanding of the feature mapping is needed before applying feature variance to the training data population. For example, it is clear that the projected location of the illuminated spot (i.e. centroid data) is embedded with information about the attitude. Corrupting the training data with large translational perturbations could lead to an ambiguous, and therefore meaningless, representation of the feature space.

As mentioned, the proposed data augmentation process follows the application of a number of filters to the pre-processed images. The data augmentation is applied through the MATLAB environment. First, the input image is duplicated for processing and shot-noise is applied through a custom function call. In addition, the read-noise is applied through the same function directly after. Finally, we add upper thresholding to the image to emulate pixel intensity saturation. The intention behind these steps is to emulate realistic uncertainties that would affect the sensor during in-flight operations. Furthermore, the data augmentation process results in an expanded training dataset with greater feature variability for improved model generalization and robustness.

\begin{figure*}
\centering
\begin{subfigure}{.32\textwidth}
    \centering
    \includegraphics[width=\linewidth]{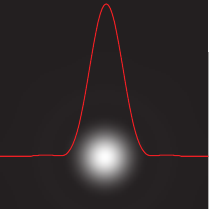}  
    \caption{Unaugmented}
    \label{fig:a}
\end{subfigure}
\begin{subfigure}{.32\textwidth}
    \centering
    \includegraphics[width=\linewidth]{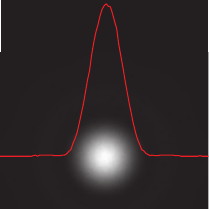}  
    \caption{Noise}
    \label{fig:b}
\end{subfigure}
\begin{subfigure}{.32\textwidth}
    \centering
    \includegraphics[width=\linewidth]{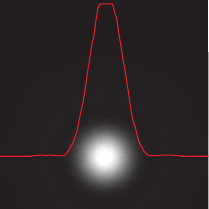}  
    \caption{Noise + Thresholding}
    \label{fig:c}
\end{subfigure}
\caption{Data augmentation with image intensity profile superimposed (a) Original (b) Noise (c) Noise + Thresholding.}
\label{fig:dataaug}
\end{figure*}

We provide access to the calibration image dataset on \href{https://doi.org/10.5281/zenodo.15778886}{Zenodo} \cite{Herman2025d}. Furthermore, the full list of associated software and data repositories is detailed in Appendix \ref{sec:appendixa}.

\section{Experiments}\label{sec:methods}

In this section, we introduce the development of the proposed network architecture and the associated training process.

\begin{figure*}[ht]
\noindent\rule{\textwidth}{1.25pt} 
\centerline{\includegraphics[width=\textwidth]{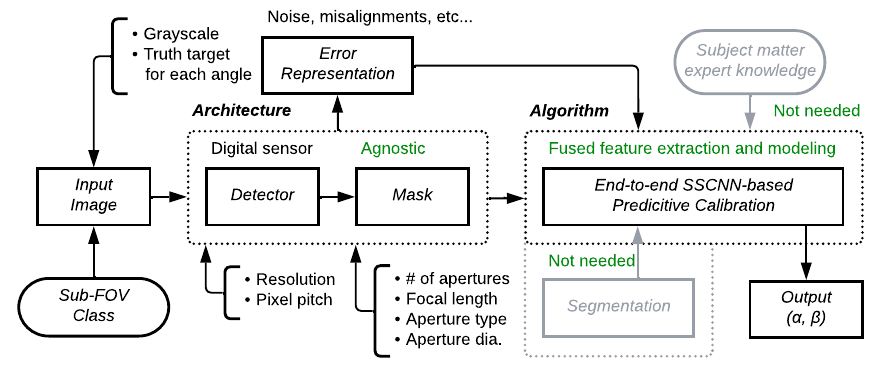}}
\noindent\rule{\textwidth}{1.25pt} 
\caption{Our proposed digital sun sensor calibration process flow. We note the advantages of our proposed method compare to the traditional calibration process in green text.}
\label{fig:propcalflow}
\end{figure*}

\subsection{Proposed network architecture}

\begin{figure*}[ht]
\noindent\rule{\textwidth}{1.25pt} 
	\centering
		\includegraphics[width=\textwidth]{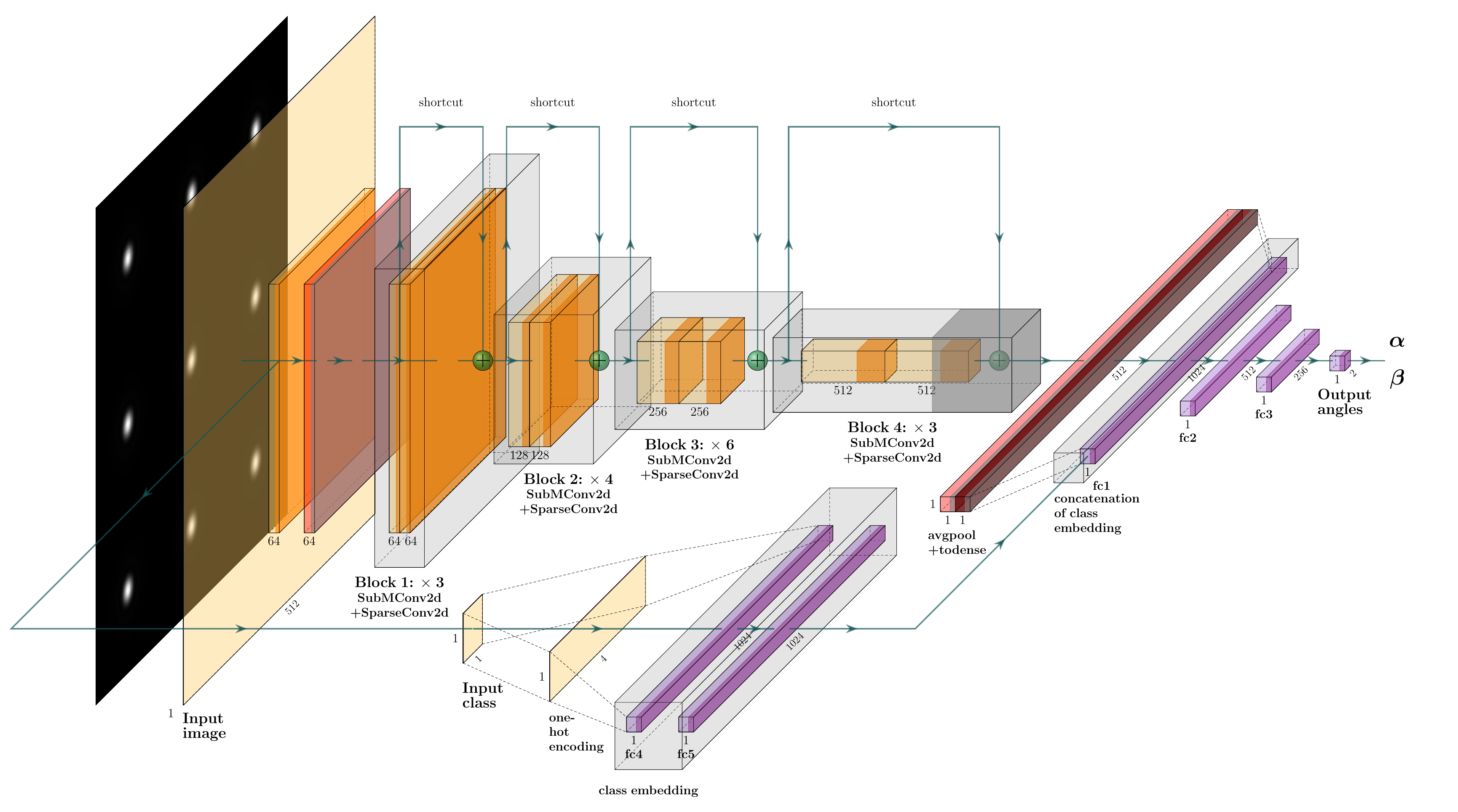}
\noindent\rule{\textwidth}{1.25pt} 
	\caption{Our proposed sparse CNN network architecture (DSS-Net). It is based on a ResNet-34 backbone with sparse submanifold convolutional layers (SSCNN) to handle sparsity.}
	\label{fig:sparseresnet}
\end{figure*}

We present a convolutional neural network (CNN) architecture based on the ResNet-34 backbone \cite{He2015}, designed to regress the two solar incident angles from a digital sun sensor’s illumination pattern and a sub-field-of-view (sub-FOV) class. ResNet gained prominence as the winner of the ILSVRC 2015 competition in image classification, detection, and localization. The depth of ResNet networks ranges from 18 to 152 layers. Its primary advantage lies in the ability to train deeper networks by learning residual functions instead of directly learning unreferenced functions. In standard networks, increasing depth often leads to the vanishing gradient problem.

\begin{figure}[ht]
	\centering
        \includegraphics[width=0.5\textwidth]{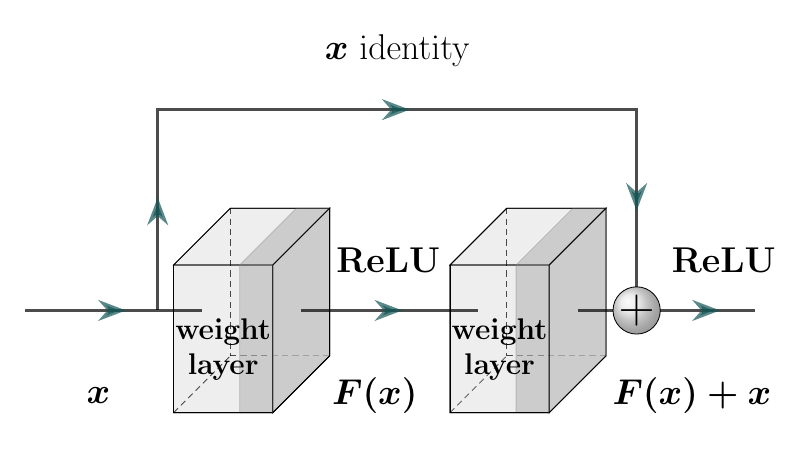}
	\caption{Residual block with shortcut connection to solve the problem of vanishing gradients.}
	\label{fig:shortcut}
\end{figure}

To address this, ResNet introduced residual blocks with shortcut connections, as illustrated in Figure \ref{fig:shortcut}. These shortcut connections perform identity mapping, where the inputs $x$ are directly added to the outputs of stacked weight layers. As a result, the weight layers learn an approximate residual mapping, which mitigates the vanishing gradient issue by allowing gradients to flow more effectively through the network layers. Additionally, despite its depth, even ResNet-152 has lower computational complexity than VGG-16 because shortcut connections do not increase the number of parameters or the computational burden of the network.


\textbf{Sparse convolutional neural networks.}
Convolutional neural networks are state-of-the-art architectures for processing image-like data, and have shown promising results in space sensor state estimation. While deep neural network calibration techniques have been successfully applied to analog sun sensors, to the best of our knowledge, CNNs have not yet been applied to digital sun sensors. Analog sensors benefit from data that is continuous, dense, and one-dimensional.

In contrast, illumination images from digital sun sensors introduce a set of unique challenges:


\begin{itemize}
    \item \textbf{Discrete:} The sensor response is composed of a discretized set of pixels instead of a continuous set.
    \item \textbf{Sparse:} The meaningful features in the detector space make up a very small portion of the total image.
    \item \textbf{Multi-dimensional:} The digital sensor data takes the form of a multi-dimensional image rather than a one-dimensional set of voltages.
\end{itemize}

Traditionally, the calibration task for digital sensors is accomplished in two stages with a feature extractor and a model to represent the specific sensor architecture. This method is computationally inefficient, limited by architecture, and requires subject matter expert knowledge to develop. Furthermore, the feature extraction is generally done through centroiding and suffers for a sub-pixel accuracy floor that is difficult to conventionally surpass.

Since dense CNNs operate on all pixels in the feature space, this can lead to computational burdens when operating on sparse data. Furthermore, dense networks struggle to learn from the sparse feature spaces. As such, the challenges due to the sparsity inherent in digital sun sensor illumination patterns necessitate the usage of sparse convolutional neural networks. Hence, we propose to convert the illumination patterns into sparse tensors to leverage sparse and sparse submanifold convolutional neural networks (SSCNNs).

\textbf{We outline the advantages of our proposed method:}

\begin{enumerate}[nolistsep]
    \item \textbf{Data-driven decision making:} The proposed methodology leverages data-driven predictive calibration to formulate the model, its parameters, and the uncertainty characterization of the response. In this case, only sensor images, a sub-FOV class, and the two sun angle targets are required to fully construct a calibration mapping. This approach requires less manual model formulation to achieve digital sun sensor calibration, while also enabling a higher-accuracy sensor mapping due to lower model uncertainty.
    \item \textbf{Separate FE and model not required:} Traditionally, the calibration for digital sensors is a two-stage process, with a feature extractor and a model representation for a given sensor architecture. We propose an end-to-end fused feature extraction and model representation framework for digital sun sensor predictive calibration. By employing this approach, the calibration process is greatly simplified with a single inference model.
    \item \textbf{No segmentation required:} The implementation of SSCNNs means that only non-sparse pixels are operated on, thereby greatly reducing computational overhead and removing the burden of segmentation methods.
    \item \textbf{Richer feature extraction:} Since feature extraction is achieved through sparse convolutional layers rather than traditional centroiding, we are able to extract more information from features to improve the sub-pixel accuracy floor that is conventionally difficult to surpass. Furthermore, we are even able to leverage traditionally detrimental features such as diffraction rings to improve the estimate.
    \item \textbf{Mask agnostic:} Our approach works for any mask configuration, as long as it corresponds to an unambiguous illumination pattern for a given target angle.
    \item \textbf{No denoising required:} CNNs are able to implicitly learn to ignore noise by training on noisy sensor data. This allows for much more accurate feature extraction, since traditional centroiding is limited by the noise floor.
    \item \textbf{Correlation for sub-FOVs via classes:} This method is able to account for sub-FOVs to achieve multiplexing capabilities from class inputs. Each sub-FOV can be labeled as a class to improve sub-region accuracy and extend the total FOV. Classes can be generated either manually with the images or from a classifier in front of the regression network.
\end{enumerate}

The sparsity of the detector space will no longer be a problem, as the number of computations only depends on the number and size of the illumination spots incident on the detector. This improved computational efficiency enables the use of deep learning on edge devices for real-time satellite operations. In addition, since the feature extraction and model representation processes are fused in this approach, the methodology is amenable to any aperture shape and mask pattern. As mentioned, our approach exploits the image sparsity by replacing traditional convolutional layers with a combination of sparse and sparse submanifold convolutions. Unlike traditional convolutions which extract features by traversing a kernel over all of the input data, sparse submanifold convolutions only operate on non-zero pixel values.

This architecture feature solves the challenges of training dense CNNs on sparse digital sun sensor images, since most of the learning computations are wasted on zero entries. Moreover, submanifold convolutions have the added benefit of maintaining the original data coordinates, thereby retaining the embedded feature localization information in each point. Regular sparse convolutions have the downside of dilating all sparse features after each layer. This effect causes the features to spread and make the data progressively less sparse as it passes through very deep networks \cite{Chen2022}. The increase in computations from reduced sparsity ultimately decreases the network efficiency. For this application, we limit the use of regular sparse convolutions to only the downsampling layers.

\begin{figure}[ht]
	\centering
        \includegraphics[width=0.5\textwidth]{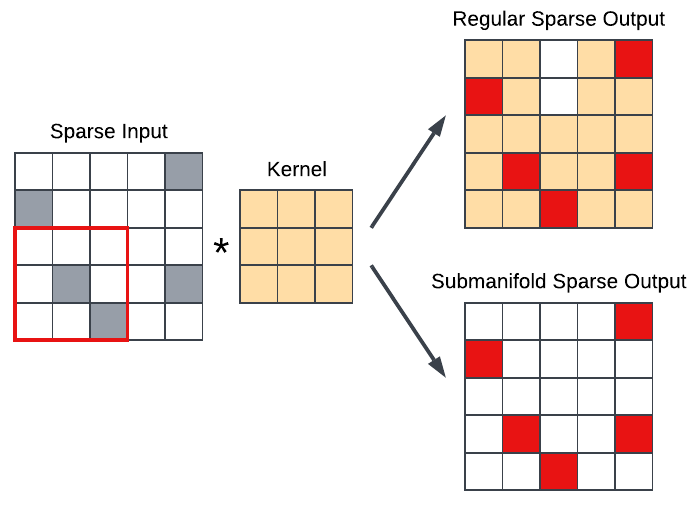}
	\caption{Regular sparse convolutions dilate the sparse features after each layer, thereby reducing the inherent sparsity of the data. However, sparse submanifold convolutions only operate directly on the sparse locations.}
	\label{fig:sparseconv}
\end{figure}

Conversely, sparse submanifold convolutions preserve the output sparsity by matching the output feature coordinates to those of the input. Furthermore, the SSCNNs differ from SCNNs because they only operate on submanifolds, which is a lower dimension subset of the sparse input data. We show the difference in behavior between SSCNNs and SCNNs in Figure \ref{fig:sparseconv}.

\textbf{Proposed SSCNN architecture.} \label{sec:propmod} Since the original dense ResNet architecture cannot be directly applied to this problem, we propose a modified ResNet34-based regression SSCNN to solve for the two sun angles, as shown in Figure \ref{fig:sparseresnet}. The network requires two inputs, which include one $512\times512$ single-channel illumination image and one class label. The output to the network is two-axis sun angle estimates in radians. Our proposed SSCNN is composed of five blocks, including one regression layer and four residual blocks with 3,4,6,3 layer configurations, respectively. Furthermore, a class embedding block is used to concatenate the class encoding with the flattened convolutional output.

We develop the network from scratch using the PyTorch deep learning framework \cite{Paszke2017} and the Spatially Sparse Convolution Library (SpConv) \cite{Yan2022} for the sparse convolutions. The work by \citet{Lathuiliere2020} was used as a guide to build the network architecture and tune the hyperparameters for this study. Since this process is highly dependent on the dataset characteristics, we also made our own adjustments to hone in on the best estimates possible.

We initialize the network by converting input image batches from dense tensor types to \texttt{SparseConvTensor} types using the Spconv \texttt{from\_dense} function. The first layer is a strided sparse convolutional layer, which downsamples the input using a kernel size of $63 \times 7 \times 7$, a stride of 2, and padding of 3. This is followed by a ReLU activation and a sparse max-pooling layer. In line with the findings of \citet{Lathuiliere2020}, we found that batch normalization (BN) does not enhance regression estimates, so it is excluded from our implementation. The first four blocks of the network consist of multiple sparse residual blocks, where sparse submanifold convolutions with same padding are used for the primary layers, and strided sparse convolutions are used only in the downsampling layers.


The input is then processed through four primary blocks, each comprising a specific number of residual blocks: three in the first block, four in the second, six in the third, and three in the fourth and final block. The sparse convolutional layers in these blocks use kernels of sizes $64 \times 3 \times 3$, $128 \times 3 \times 3$, $256 \times 3 \times 3$, and $512 \times 3 \times 3$, respectively. Each primary block concludes with a strided sparse convolutional layer for downsampling, which has a kernel size of 1 and a stride of 2. Additionally, downsampling is applied in the layer preceding the final skip connection of each primary block to ensure the residuals being added have matching dimensions. Notably, we use sparse convolutions instead of sparse submanifold convolutions in the downsampling layers, as only sparse convolutions can perform downsampling. This distinction arises because sparse submanifold convolutions maintain the original sparsity pattern by preserving spatial dimensions.


We complete the network architecture by converting the SpConv sparse tensor into a dense tensor and reformatting it to the NCHW layout. In this format, $N$ represents the batch size, $C$ the number of channels, $H$ the height, and $W$ the width of the data. The dense tensor is then processed through average pooling with a kernel size of 7 and a stride of 1.

Before constructing the network’s linear layers, we encode the class input using one-hot encoding to convey sub-FOV information to the network. This step improves the final sun angle estimates by constraining the prediction within the relevant region. The class encoding is passed through a class embedding block, which consists of two linear layers with ReLU activations. The output of this block is concatenated with the flattened convolutional output. This concatenation enables the sub-FOV information to enhance the features extracted by the convolutional layers, thereby improving the accuracy of the estimates by incorporating sub-FOV-specific context not captured by the convolutional layers alone.


The network architecture concludes with a series of five linear layers, each followed by a ReLU activation, and ends with a \texttt{Tanh} operator. Including the \texttt{Tanh} function at the network’s output improves estimation accuracy due to its zero-centered output, range of [-1, 1], non-linearity, and stronger gradients near zero. The first four linear layers produce outputs with feature sizes of 1024, 512, 256, and 256, respectively. Instead of using a softmax layer for classification, the final linear layer employs the identity function to facilitate regression. As a result, the network outputs a real-valued $2 \times 1$ matrix representing the sun angles $(\alpha, \beta)$ in radians.


Here, we provide additional details about the architecture implementation within the sparse convolution framework. We utilized the SpConv v2.3.6 library \cite{Yan2018, Yan2022} with CUDA v12.0 to implement the \texttt{SparseConv2d} and \texttt{SubMConv2d} convolutional layers. Specifically, the mask implicit general matrix-matrix multiplication (GEMM) algorithm was chosen for its ability to enhance computational efficiency on GPU architectures.

SpConv v2.x re-implements the CUDA Templates for Linear Algebra Subroutines and Solvers (CUTLASS) \cite{Kerr2022} dataflow, optimized for sparse workloads, to enable faster sparse convolution operations \cite{Tang2023}. A distinctive feature of SpConv is its ability to pass index keys to subsequent sparse submanifold convolutions with the same spatial structure, further accelerating computations.

The SSCNN training and testing code is available on \href{https://doi.org/10.5281/zenodo.16579535}{Zenodo} \cite{Herman2025f}. The trained SSCNN models are available on \href{https://doi.org/10.5281/zenodo.15778990}{Zenodo} \cite{Herman2025e}. Moreover, the full list of associated software and data repositories is detailed in Appendix \ref{sec:appendixa}.


\subsection{Training details}

\begin{figure*}[ht]
	\centering
		\includegraphics[width=\textwidth]{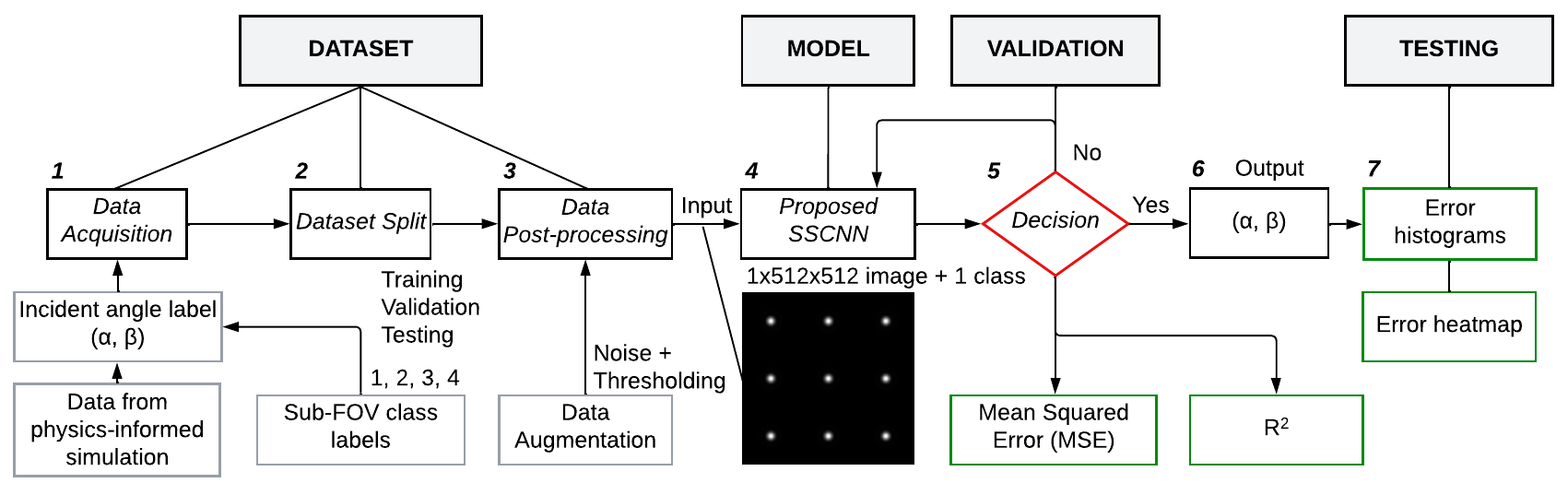}
	\caption{Methodology flow diagram.}
	\label{fig:trainflow}
\end{figure*}

In this section we describe the full end-to-end training process. The training workflow of the calibration network is presented in Figure \ref{fig:trainflow}. This process takes the form of five steps including data acquisition, dataset split, data post-processing, development of the proposed model architecture, and model training and validation.

\hypertarget{step:one}{\textbf{Step 1: Data Aquisition.}} The first step in the training process is data acquisition, where the synthetic data is generated from physics-informed simulations and assigned targets and sub-FOV classes. We detail out the data acquisition process in Section \ref{sec:daq}.

\hypertarget{step:two}{\textbf{Step 2: Dataset Split.}} The second step in the training process is the dataset split. We divide the data into a training set, a validation set, and a testing set. The training set is used for supervised  learning to train the model weights. The validation set is used to verify the model generalization and tune hyperparameters. Finally, the testing set is used to assess the model performance on augmented and non-augmented data for the two mask configurations.

We present the dataset split used for this study in Table \ref{tab:alldata}. The data was split into the three sets at a ratio of 8:1:1, respectively. Moreover, we assigned data to the three sets at random, while ensuring that the aforementioned ratio is maintained. The total dataset size is 789,507 for a single mask configuration with augmented data. Furthermore, since there are three augmentation image types (original,noise, thresholding), the size of the non-augmented dataset is 263,169. However, we found during training that around 65,536 images were sufficient for good convergence during training. This yields a proposed sampling density of around $65,536/(512\times512)=1/4$, which greatly reduces the amount of data needed to be generated via simulation or ground testing to use this method.

\begin{table}[htb]
\caption{Training data for all mask configurations.}
\label{tab:alldata}
\centering
\setlist[itemize]{wide, leftmargin=*, noitemsep, after=\vspace*{-\topsep}}
\setlength{\extrarowheight}{3pt}
\setlength{\tabcolsep}{3pt}
\setlength\ct{\dimexpr .167\textwidth -4\tabcolsep}
\noindent\begin{tabular}{>{\compress\rowmac}p{\ct}|>{\compress\rowmac}P{\ct}|>{\compress\rowmac}P{\ct}<{\clearrow}}
\Xhline{1.25pt}
\textbf{Part} & \textbf{Proportion} & \textbf{Quantity}\\
\hline
Total & 1 & 789,507\\
Training set & 0.8 & 631,605\\
Validation set & 0.1 & 78,951\\
Testing set & 0.1 & 78,951\\
\hline
All masks & 2 & 1,579,014\\
\Xhline{1.25pt}
\end{tabular}
\end{table}

\hypertarget{step:three}{\textbf{Step 3: Data Post-processing.}} Next, in the third step in the training process involves augmenting the data with post-processing techniques. For this study, we added realistic sensor noise and thresholding to the images. An in-depth discussion on this process in presented in Section \ref{sec:daug}.

\hypertarget{step:four}{\textbf{Step 4: Proposed model architecture.}} This data ($1\times512\times512$ image + 1 class), is then passed to the proposed SSCNN model in \hyperlink{step:four}{Step 4}, as discussed in Section \ref{sec:propmod}.

\hypertarget{step:five}{\textbf{Step 5: Model training.}} An important part of the training process is the selection and tuning of hyperparameters to ensure the model converges to the most optimal solution possible during training. Therefore, for \hyperlink{step:five}{Step 5} we train the model and tune hyperparameters, where the work by \citet{Lathuiliere2020} was once again used as a guide for this study along with independent testing for each mask configuration. An overview of the hyperparameters used for each mask and data configuration is presented in Table \ref{tab:hyperparameters}.

\begin{table*}[htb]
\caption{Training hyperparameters for each mask and data configurations.}
\label{tab:hyperparameters}
\centering
\setlist[itemize]{wide, leftmargin=*, noitemsep, after=\vspace*{-\topsep}}
\setlength{\extrarowheight}{3pt}
\setlength{\tabcolsep}{3pt}
\setlength\ct{\dimexpr .25\textwidth -2\tabcolsep}
\noindent\begin{tabular}{>{\compress\rowmac}p{0.75\ct}?>{\compress\rowmac}P{1.1\ct}>{\compress\rowmac}P{0.5\ct}|>{\compress\rowmac}P{1.1\ct}>{\compress\rowmac}P{0.5\ct}<{\clearrow}}
\Xhline{1.25pt}
\textbf{Hyperparameter} & \textbf{Single-aperture} & \textbf{Augmented} & \textbf{Multi-aperture} & \textbf{Augmented}\\
\hline
Loss function & MSE & - & MSE & -\\
Loss scaling & 100 & - & 100 & -\\
\hline
Optimizer & Adamax & - & Adamax & -\\
Learning rate & $1\mathrm{e}-3$ & - & $3\mathrm{e}-3$ & -\\
Weight decay & 0 & - & 0 & -\\
\hline
Epochs & 100 & - & 100 & -\\
Batch size & 1024 & - & 325 & -\\
\hline
Scheduler & Cosine Annealing w/ WR & - & Cosine Annealing w/ WR & -\\
$T_{0}$ & 5 & - & 4 & -\\
$T_{mult}$ & 1 & - & 1 & -\\
Decay & 0.7 & - & 0.4 & -\\
Warmup scheduler & None & - & Exponential & -\\
$T_{warmup}$ & None & - & 25 & -\\
\hline
Early stopping & True & - & True & -\\
Patience & 4 & 6 & 3 & 5\\
$\Delta_{min}$ & 0 & - & 0 & -\\
\Xhline{1.25pt}
\end{tabular}
\end{table*}

\textbf{Loss function.} As shown in \hyperlink{step:five}{Step 5} of Figure \ref{fig:trainflow}, a decision point is hit during training, in which the model converges to the target values to minimize the loss function. For this work, the loss function of mean squared error (MSE) is selected for the regression task along with the coefficient of determination ($R^{2}$) for model fitness. We also tried the loss functions of mean absolute error (MAE) and Huber loss, however we found that MSE proved to be the most reliable. This is backed up by the results found by \citet{Lathuiliere2020}. We want the MSE value to be as low as possible and the $R^{2}$ to approach 1.

The formulation of the MSE is presented in Equation \ref{eq:mse}:

\begin{equation}\label{eq:mse}
MSE_{V}=\frac{1}{M}\sum_{m=1}^{M}\left ( y_{m}^{V}-\hat{y}_{m}^{V} \right )^{2}
\end{equation}

where, $M$ is the total number of training samples, $m$ is the m-th iterated sample of the total $M$ samples, $y_{m}^{V}$ is the observed state estimate value, and $\hat{y}_{m}^{V}$ is predicted state estimate value. In addition, $MSE_{V}$ is the MSE for the given individual state estimate of interest.

In addition, we also evaluate the coefficient of determination ($R^{2}$) to assess the model fitness to the data during training. The formulation of the $R^{2}$ value is presented in Equation \ref{eq:r2}:

\begin{equation} \label{eq:r2}
R^{2}=1-\frac{\sum_{m=1}^{M}\left ( \hat{y}_{m}-y_{m} \right )^{2}}{\sum_{m=1}^{M}\left ( \bar{y}_{m}-y_{m} \right )^{2}}
\end{equation}

where, $R^{2}$ is the coefficient of determination, $M$ is the total number of training samples, $m$ is the m-th iterated sample of the total $M$ samples, $\hat{y}_{m}$ is the predicted value of $y_{m}$, ${y}_{m}$ is the m-th value of the state to be predicted, and $\bar{y}_{m}$ is the mean value of the m-th state sample. In addition, $\left ( \hat{y}_{m}-y_{m} \right )^{2}$ is the residual sum of squares and $\left ( \bar{y}_{m}-y_{m} \right )^{2}$ is the total sum of squares.

\textbf{Optimizer.} We selected Adamax as the optimizer of choice since it is more stable for sparse gradients. Adamx is a variant of the Adam optimizer based on the infinity norm. In addition, we also tested the Adam and AdamW optimizers, however neither showed noticeable improvements. We chose the maximum number of epochs as 100, however this was never reached due to the early stopping criteria.

\textbf{Batch size.} The batch sizes for each mask configuration were selected as the largest possible given the VRAM available. Following the recommendations by \citet{Lathuiliere2020}, it was found that while having no affect on accuracy, higher batch sizes led to better training performance. Furthermore, through independent testing we found that larger batch sizes smoothed the gradients, thereby improving convergence. As such, we selected a batch size of 1024 for the single-aperture configuration and a batch size of the 325 for the multi-aperture configuration. As expected, the more features present in the image space, the more memory required for training the model. Nevertheless, the use of sparse networks greatly reduced the total memory required for training compared to a dense model variant.

\textbf{Scheduler.} For the learning rate scheduler, we selected a variable learning rate for the training process. Specifically, cosine annealing with warm restarts and decay was implemented. This is a modified variant of the \\\texttt{CosineAnnealingWarmRestarts} to add a decay factor. We used a variable learning rate since we noticed a high sensitivity of the learning rate to the convergence performance, especially at early epochs. Furthermore, decay was added since a smaller learning rate was required over epochs to avoided overshooting the optima.

\textbf{Early stopping.} To end each training run we implemented early stopping conditions. For the single-aperture configuration without augmentation we used a patience of 4, while we used a patience of 6 with augmentation. Furthermore, for the multi-aperture configuration without augmentation we used a patience of 3, while we used a patience of 5 with augmentation. These patience values were decided via trial and error for each case to avoid overfitting the data.

\begin{figure*}[H]
    \centering
    \begin{subfigure}[b]{\textwidth}
        \centering
        \includegraphics[width=\textwidth]{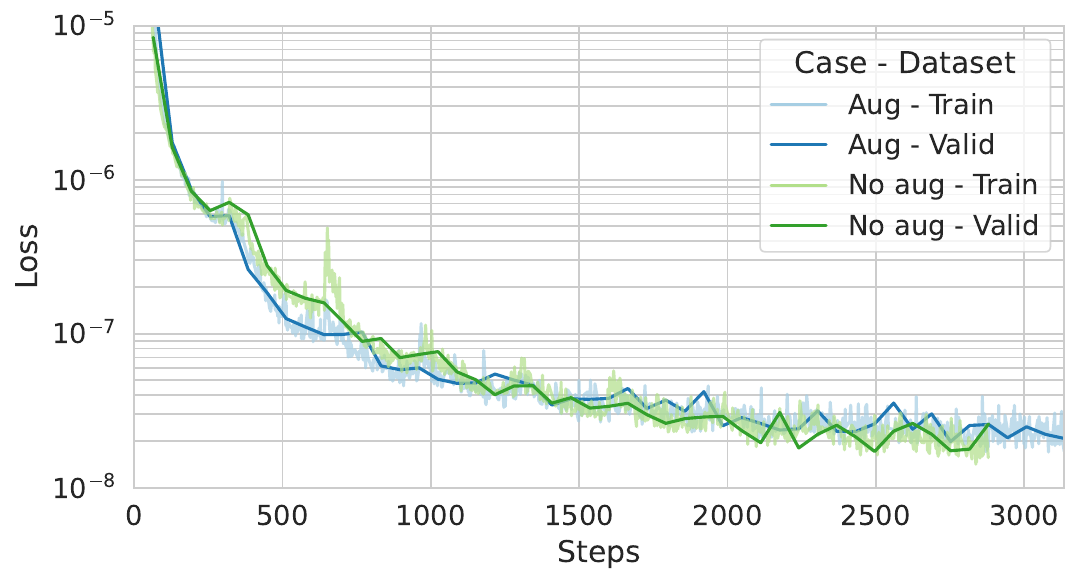}
        \caption{MSE loss of single-aperture configuration for augmented and non-augmented models with training and validation data over the full training process.}
        \label{fig:sa_loss}
    \end{subfigure}
    
    \vspace{0.25cm} 
    
    \begin{subfigure}[b]{0.49\textwidth}
        \centering
        \includegraphics[width=\textwidth]{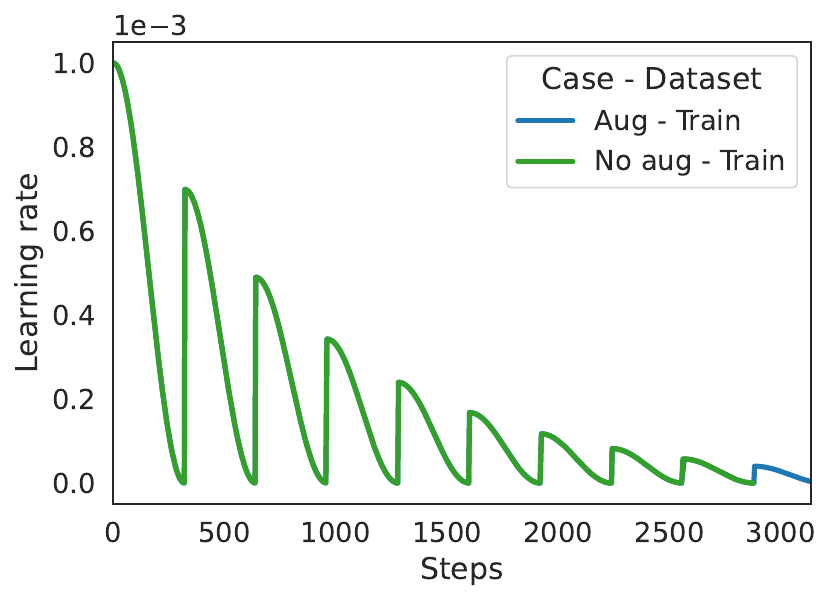}
        \caption{Learning rate of single-aperture configuration for augmented and non-augmented models with training data over the full training process.}
        \label{fig:sa_lr}
    \end{subfigure}
    \hfill
    \begin{subfigure}[b]{0.49\textwidth}
        \centering
        \includegraphics[width=\textwidth]{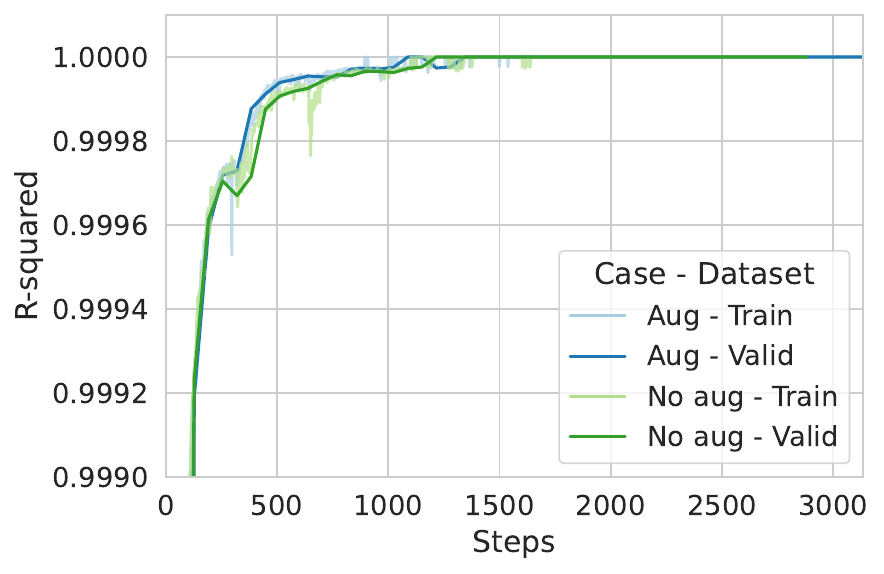}
        \caption{R-squared of single-aperture configuration for augmented and non-augmented models with training and validation data over the full training process.}
        \label{fig:sa_r2}
    \end{subfigure}
    
    \caption{Training metrics for the single-aperture configuration for augmented and non-augmented models over the full training process.}
    \label{fig:sa_train_overall}
\end{figure*}

\textbf{Single-aperture training.} Next, we will outline the single-aperture configuration training process and associated hyperparameters. The Adamax optimizer is used with an initial learning rate of $1\mathrm{e}-3$ with a weight decay of 0. We use a learning rate scheduler of cosine annealing with warm restarts and decay. The cosine period $T_{0}$ is 5, the cosine period multiplier $T_{mult}$ is 1, the decay factor is 0.7, and no warmup schedule is used. The learning rate for the entire single-aperture model training process is presented in Figure \ref{fig:sa_lr}. The augmented model is blue, whereas the non-augmented model is green.

The loss function of MSE is used for training the single-aperture model. We use a loss scaling factor of 100 to scale the gradients and thereby improve convergence, since the target values are very small. The loss for single-aperture model training process is presented in Figure \ref{fig:sa_loss}. The augmented model is blue, whereas the non-augmented model is green. Furthermore, the training data is a lighter shade and the validation data is darker shade for each respective color.

The loss starts off by rapidly decreasing until around 1000 steps, whereafter it slowly converges to around 3000 steps. It is of interest that the augmented and non-augmented models follow nearly identical convergence trends, which suggests that the feature pattern has an outsized influence on the convergence behavior. We believe that the minimum converged to is very close to the loss floor for this model and input combination. This is due to the loss wall we appeared to hit after plateauing in the loss landscape.

The R-squared function for the single-aperture model shows good convergence to the value of 1. We present the R-squared for the single-aperture model training process in Figure \ref{fig:sa_r2}. The augmented model is blue, whereas the non-augmented model is green. Furthermore, the training data is a lighter shade and the validation data is darker shade for each respective color. It is clear that the correlation coefficient dramatically increases until around 1000 steps, and then slowly converges toward 1 until the stop condition.

\textbf{Multi-aperture training.} Finally, we additionally outline the multi-aperture configuration training process and associated hyperparameters. The Adamax optimizer is used with an initial learning rate of $3\mathrm{e}-3$ with a weight decay of 0. We use a learning rate scheduler of cosine annealing with warm restarts and decay. The cosine period $T_{0}$ is 4, the cosine period multiplier $T_{mult}$ is 1, the decay factor is 0.4, and a exponential warmup is implemented with a warmup period $T_{warmup}$ of 25 steps.

We utilized a warmup scheduler using the \\\texttt{pytorch\_scheduler} library \cite{Yamamoto2024}. The warmup schedule was implemeted because the training process would diverge with an initial learning rate above $1\mathrm{e}-3$, however such a learning rate was needed to properly converge for this case. It is hypothesized that the learning rate needs to be higher for the multi-aperture cases due to the increased number of features in the image space.

\begin{figure*}[H]
    \centering
    \begin{subfigure}[b]{\textwidth}
        \centering
        \includegraphics[width=\textwidth]{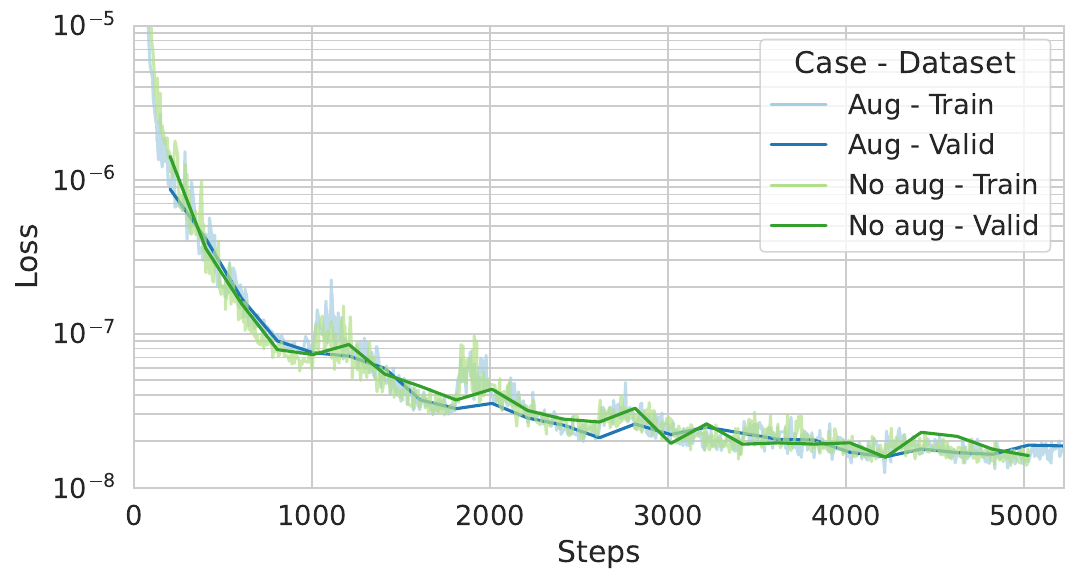}
        \caption{MSE loss of multi-aperture configuration for augmented and non-augmented models with training and validation data over the full training process.}
        \label{fig:ma_loss}
    \end{subfigure}
    
    \vspace{0.25cm} 
    
    \begin{subfigure}[b]{0.49\textwidth}
        \centering
        \includegraphics[width=\textwidth]{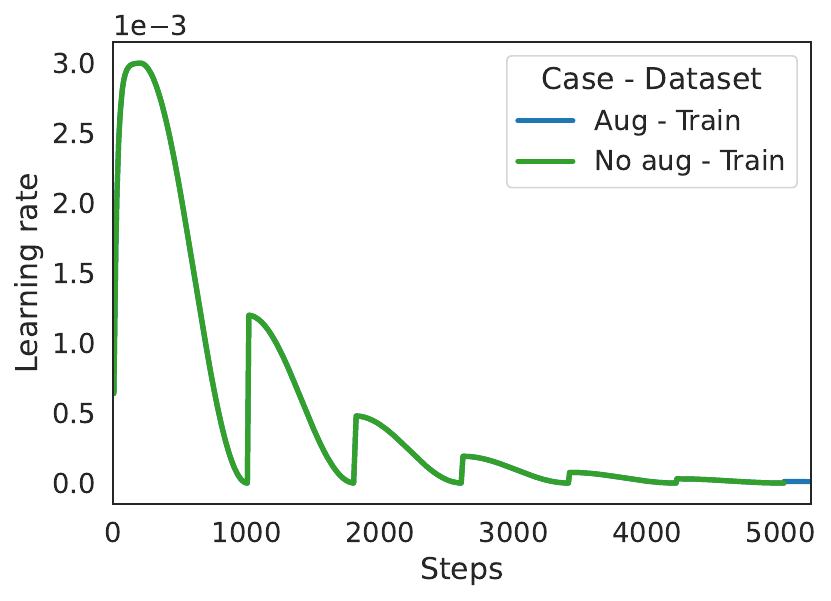}
        \caption{Learning rate of multi-aperture configuration for augmented and non-augmented models with training data over the full training process.}
        \label{fig:ma_lr}
    \end{subfigure}
    \hfill
    \begin{subfigure}[b]{0.49\textwidth}
        \centering
        \includegraphics[width=\textwidth]{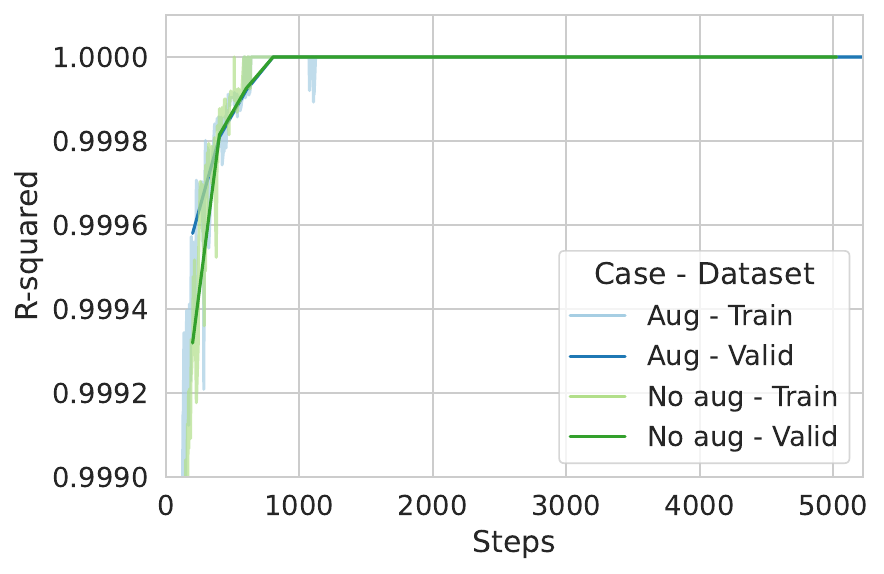}
        \caption{R-squared of multi-aperture configuration for augmented and non-augmented models with training and validation data over the full training process.}
        \label{fig:ma_r2}
    \end{subfigure}
    
    \caption{Training metrics for the multi-aperture configuration for augmented and non-augmented models over the full training process.}
    \label{fig:ma_train_overall}
\end{figure*}

The learning rate for the entire multi-aperture model training process is presented in Figure \ref{fig:ma_lr}. The augmented model is blue, whereas the non-augmented model is green.

The loss function of MSE is used for training the multi-aperture model. Just like the single-aperture case, we use a loss scaling factor of 100 to scale the gradients as the target values are very small. The loss for multi-aperture model training process is presented in Figure \ref{fig:ma_loss}. The augmented model is blue, whereas the non-augmented model is green. Furthermore, the training data is a lighter shade and the validation data is darker shade for each respective color.

The loss for the multi-aperture configuration starts off by rapidly decreasing until around 1000 steps, after which it slowly converges to around 5000 steps. Again, the augmented and non-augmented models follow nearly identical convergence trends, as expected even as the aperture count increases since the mask pattern is the same. More surprisingly, the loss value converged to is very close to that of the single-aperture configuration, even under the effects of noise and thresholding augmentation. Generally, the main advantage of multi-aperture configurations stems from the ability to effectively "average" out errors from multiple observations. However, the ability of the proposed network to implicitly ignore noise and data corruption appears to largely nullify these advantages. 

These observations suggest that the loss floor is likely most limited by pixel quantization of the feature space rather than noise errors. We note that the multi-aperture convergence is also very close to the loss floor for this model and input combination due to the loss wall encountered after plateauing in the loss landscape.

The R-squared function for the multi-aperture model converges well to the value of 1. We present the R-squared for the multi-aperture model training process in Figure \ref{fig:ma_r2}. The augmented model is blue, whereas the non-augmented model is green. Furthermore, the training data is a lighter shade and the validation data is darker shade for each respective color. The correlation coefficient quickly increases until around 1000 steps, and then gradually converges toward 1 until the early stop criteria.

\subsection{Training summary}

This concludes the training overview of the single-aperture and multi-aperture SSCNN networks under augmented and non-augmented data. We summarize the section by highlighting key findings during the training process:

\begin{enumerate}[nolistsep]
    \item \textbf{Dataset:} We find that around 65,536 training images were sufficient for good convergence during training or a proposed sampling density of around 1/4, which greatly reduces the amount of data needed to be generated via simulation or ground testing to use this method.
    \item \textbf{Model architecture:} The implementation of SSCNNs for this application is mandatory for proper convergence. The data is simply too sparse and the 
    coordinates of the active locations are too critical for localization to not leverage the SSCNN architecture. We attempted to train dense architectures with increasingly poor convergence as the number of apertures decreased (or the sparsity increased).
    \item \textbf{Learning rate:} The learning rate needs to be higher the more apertures there are due to the increased number of features in the image space. In addition, the training process is highly sensitive to the learning rate schedule and cosine annealing with warm restarts and decay is recommended.
    \item \textbf{Loss:} The final scaled loss converged to around $2\mathrm{e}-6$ for both the single and multi-aperture models. The augmented and non-augmented models follow nearly identical loss trends, even under the effects of noise and thresholding augmentation. While the main advantage of multi-aperture configurations is derived from "averaging" out errors from multiple observations, the ability of the proposed network to implicitly ignore noise and data corruption appears to largely nullify these advantages.
    \item \textbf{Loss floor:} We note that the loss floor is likely most limited by pixel quantization of the feature space rather than noise errors. Superresolution techniques could lead to even better results if implemented.
    \item \textbf{Convergence:} When under the effects of noise, the training will occasionally converge to a sub-optimal local minima and need restarted.
\end{enumerate}

\section{Results}\label{sec:result}

In order to predictively calibrate a digital sun sensor, the non-linear uncertainty of the sensor response must be accurately learned. Traditionally, this has been done by attempting to account for errors through parameterization, such as geometric and physics-informed models. However, this approach inevitably leads to model uncertainty due to the limitations of the model formulations to accurately represent the underlying physical reality of the sensor response under varying real-world conditions. To this end, the use of CNN models addresses this problem by enabling the end-to-end learning of the non-linear sensor response by mapping the input sensor images to truth targets without requiring any domain knowledge to explicitly formulate the model.

For this study, we apply the proposed SSCNN model to regress the two incident sun angles, $\alpha$ and $\beta$, for the single and multi-aperture configurations from their respective illumination pattern images. We assess the model performance by analyzing absolute error heatmaps, histograms, and predicted vs actual diagrams from the testing datasets for each configuration. The absolute error heatmaps aid in visualizing the two-axis error landscape over the full sensor FOV for each angle. Meanwhile, the error histograms clearly show the absolute error distribution as a probability density function (PDF) for each angle. This allows for easy interpretation of the error mean and spread of each angle. In addition, the predicted vs actual diagrams of the absolute errors helps to visualize how the test predictions match the ideal fit for each angle.

\subsection{Single-aperture configuration}

\begin{figure*}[H]
    \centering
    \begin{subfigure}[b]{0.95\textwidth}
        \centering
        \includegraphics[width=\textwidth]{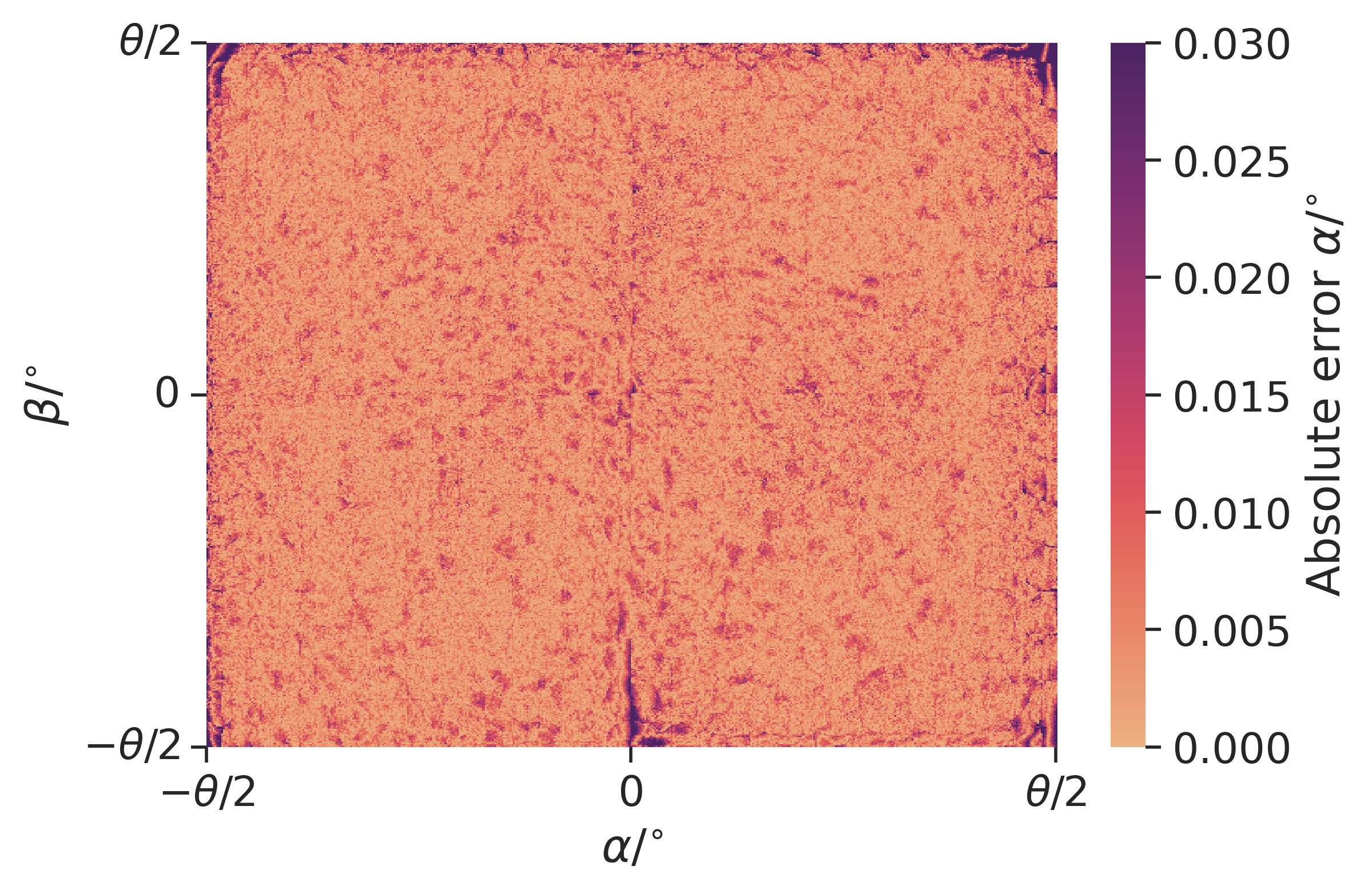}
        \caption{Absolute error heatmap for angle $\alpha$ in a single-aperture configuration with non-augmented model using the entire dataset.}
        \label{fig:sa_hma}
    \end{subfigure}
    
    \begin{subfigure}[b]{0.95\textwidth}
        \centering
        \includegraphics[width=\textwidth]{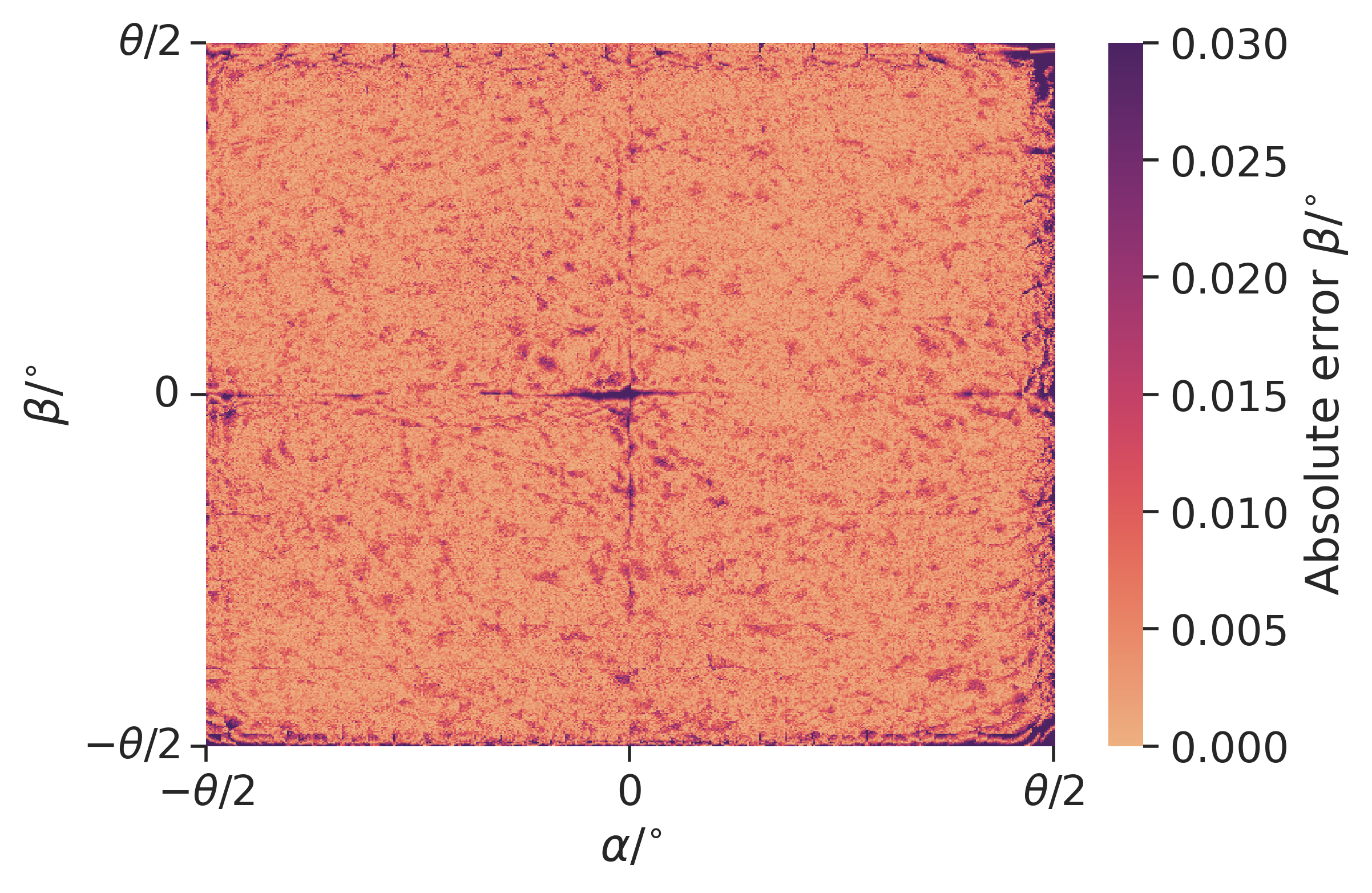}
        \caption{Absolute error heatmap for angle $\beta$ in a single-aperture configuration with non-augmented model using the entire dataset.}
        \label{fig:sa_hmb}
    \end{subfigure}
    
    \caption{Absolute error heatmaps for single-aperture configuration with non-augmented model.}
    \label{fig:sa_hm_overall}
\end{figure*}

\textbf{Absolute error heatmaps.} We begin assessing the performance of the single-aperture non-augemented model by interpreting the two-axis absolute error landscape over the full sensor FOV as illustrated by the heatmaps in Figure \ref{fig:sa_hm_overall}. The absolute error heatmap for angle $\alpha$ is shown in Figure \ref{fig:sa_hma}, while the heatmap for angle $\beta$ is shown in Figure \ref{fig:sa_hmb}. The overall error over the two-axis FOV appears to be low for both angles, where in general the error tends to be around 0.005°. While the error heatmap is mostly smooth, there are some small pockets of higher errors. This is likely due to the amount of training data used, where training on a large dataset would resolve these pockets.

Furthermore, we note the highest errors in the corners of the FOV and at the sub-FOV class boundaries. The error at the corners is expected since there is only a single aperture to observe, therefore the accuracy drops when the aperture is partially visible at the sensor edges. However, we suspect that this issue is resolved with more apertures (observable features). In addition, the errors at the class boundaries are derived from class uncertainty when traversing between sub-FOVs. In this study the sub-FOV classes are the four quadrants of the detector space, however they can be defined in any fashion that suits the mask (for example an encoded pattern). Hence, we expect to see errors due to uncertainty near the quadrant boundaries.

It appears that a single aperture as the only observed feature especially pronounces this effect of class uncertainty at the origin of the FOV. This phenomenon makes sense because as the single aperture traverses the class boundary it could be perceived as within two or more classes (sub-FOVs) simultaneously. This effect is likely to be largely resolved by using more apertures (features) or relaxing the class weights with dropout to better generalize the problem.

\begin{figure*}[H]
    \centering
    \begin{subfigure}[b]{0.49\textwidth}
        \centering
        \includegraphics[width=\linewidth]{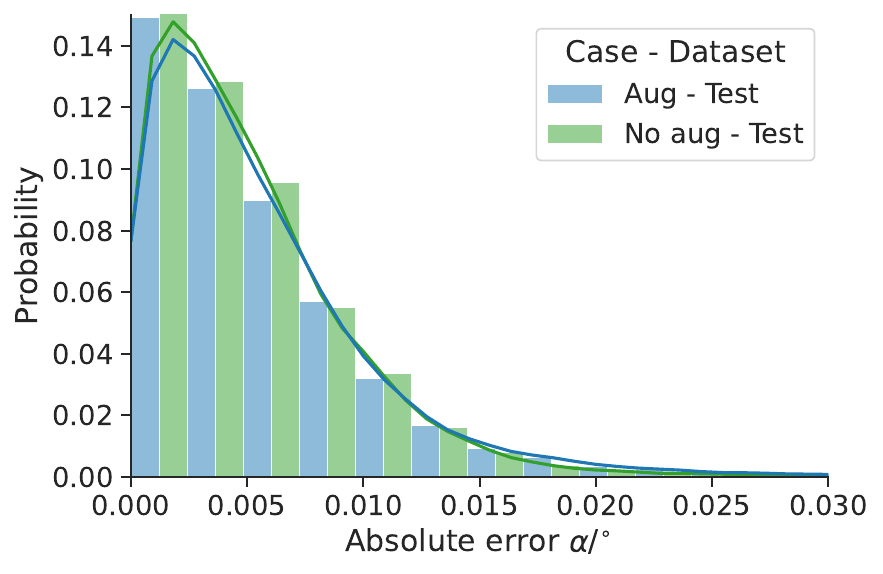}
        \caption{Absolute error histogram for angle $\alpha$ in a single-aperture configuration with augmented and non-augmented model on the testing dataset.}
        \label{fig:sa_hista}
    \end{subfigure}
    \hfill
    \begin{subfigure}[b]{0.49\textwidth}
        \centering
        \includegraphics[width=\linewidth]{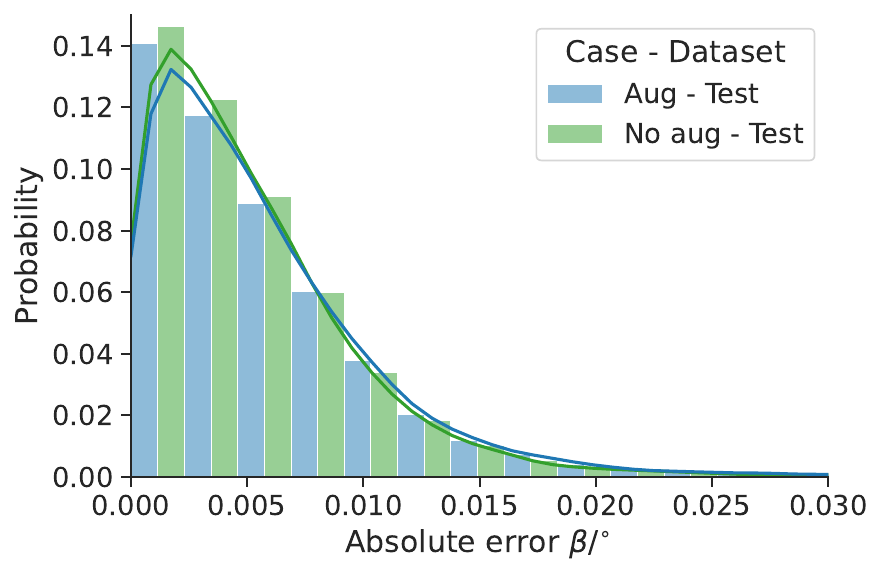}
        \caption{Absolute error histogram for angle $\beta$ in a single-aperture configuration with augmented and non-augmented model on the testing dataset.}
        \label{fig:sa_histb}
    \end{subfigure}
    
    \vspace{0.25cm} 
    
    \begin{subfigure}[b]{0.49\textwidth}
        \centering
        \includegraphics[width=\linewidth]{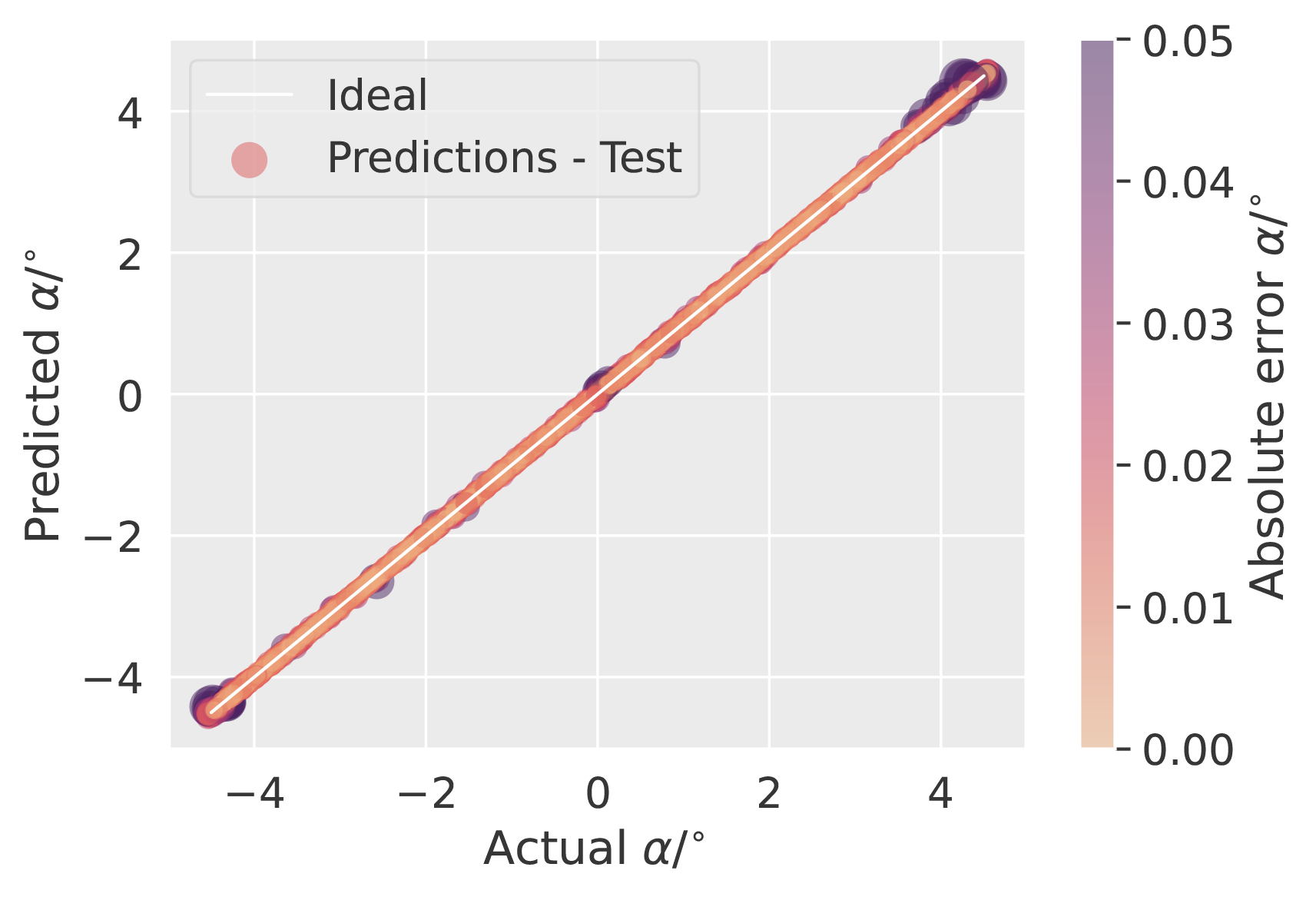}
        \caption{Predicted vs actual diagram of the absolute error for angle $\alpha$ in a single-aperture configuration with non-augmented model on the testing dataset.}
        \label{fig:sa_pvaa}
    \end{subfigure}
    \hfill
    \begin{subfigure}[b]{0.49\textwidth}
        \centering
        \includegraphics[width=\linewidth]{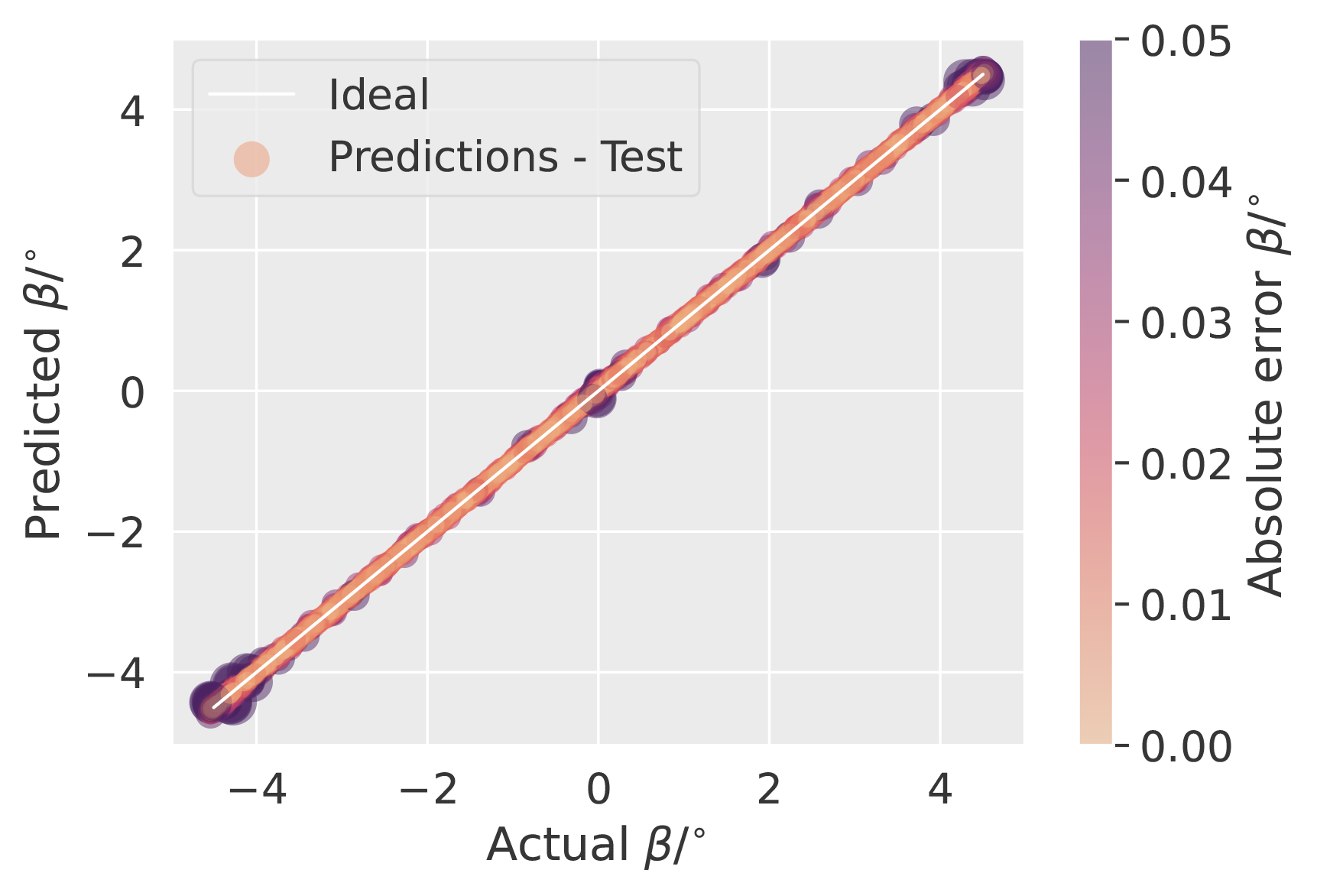}
        \caption{Predicted vs actual diagram of the absolute error for angle $\beta$ in a single-aperture configuration with non-augmented model on the testing dataset}
        \label{fig:sa_pvab}
    \end{subfigure}
    
    \caption{Absolute error metrics of the single-aperture configuration.}
    \label{fig:sa_error_overall}
\end{figure*}

\textbf{Absolute error histograms.} Next, we analyze the absolute error metrics of the single-aperture configuration in Figure \ref{fig:sa_error_overall}. First, the absolute error distributions via error histograms of the augmented and non-augmented models on the testing dataset are investigated. The absolute error histogram for angle $\alpha$ is shown in Figure \ref{fig:sa_hista}, while the error histogram for angle $\beta$ is shown in Figure \ref{fig:sa_histb}. The histograms are displayed in a "dodge" format, in which the augmented and non-augmented distributions are plotted side-by-side. As before, the augmented model is blue and the non-augmented model is green. In addition, the histograms are displayed as probability distribution functions over the error bins.

We note that both the angle estimates demonstrate similar distribution shapes, with the error spread within 0.02° and mean of about 0.005°. Furthermore, the augmented and non-augmented models display similar performance on their respective datasets. However, the non-augmented model accuracy is very slightly better than that of the augmented model. This is expected since the augmented model has to filter out the learned noise and thresholding image corruption. Nevertheless, the augmented model proves to be very robust to the learned uncertainty with a small error difference.

\textbf{Absolute error predicted vs actual.} The single-aperture non-augmented model test predictions are assessed using predicted vs actual diagrams in Figure \ref{fig:sa_error_overall}. The predicted vs actual diagram of the absolute error for angle $\alpha$ is presented in Figure \ref{fig:sa_pvaa}, while the predicted vs actual diagram of the absolute error for angle $\beta$ is presented in Figure \ref{fig:sa_pvab}. The white line portrays the ideal prediction trend, while the model test predictions are plotted as a scatter plot. The size and colormap of the scatter points vary with the magnitude of the absolute error. Overall we see a good fit of the testing data to the actual targets, with most of the errors residing at the FOV corners and the origin. This error behavior follows the same trends as seen in the heatmap previously presented.

\subsection{Multi-aperture configuration}

\begin{figure*}[H]
    \centering
    \begin{subfigure}[b]{0.95\textwidth}
        \centering
        \includegraphics[width=\textwidth]{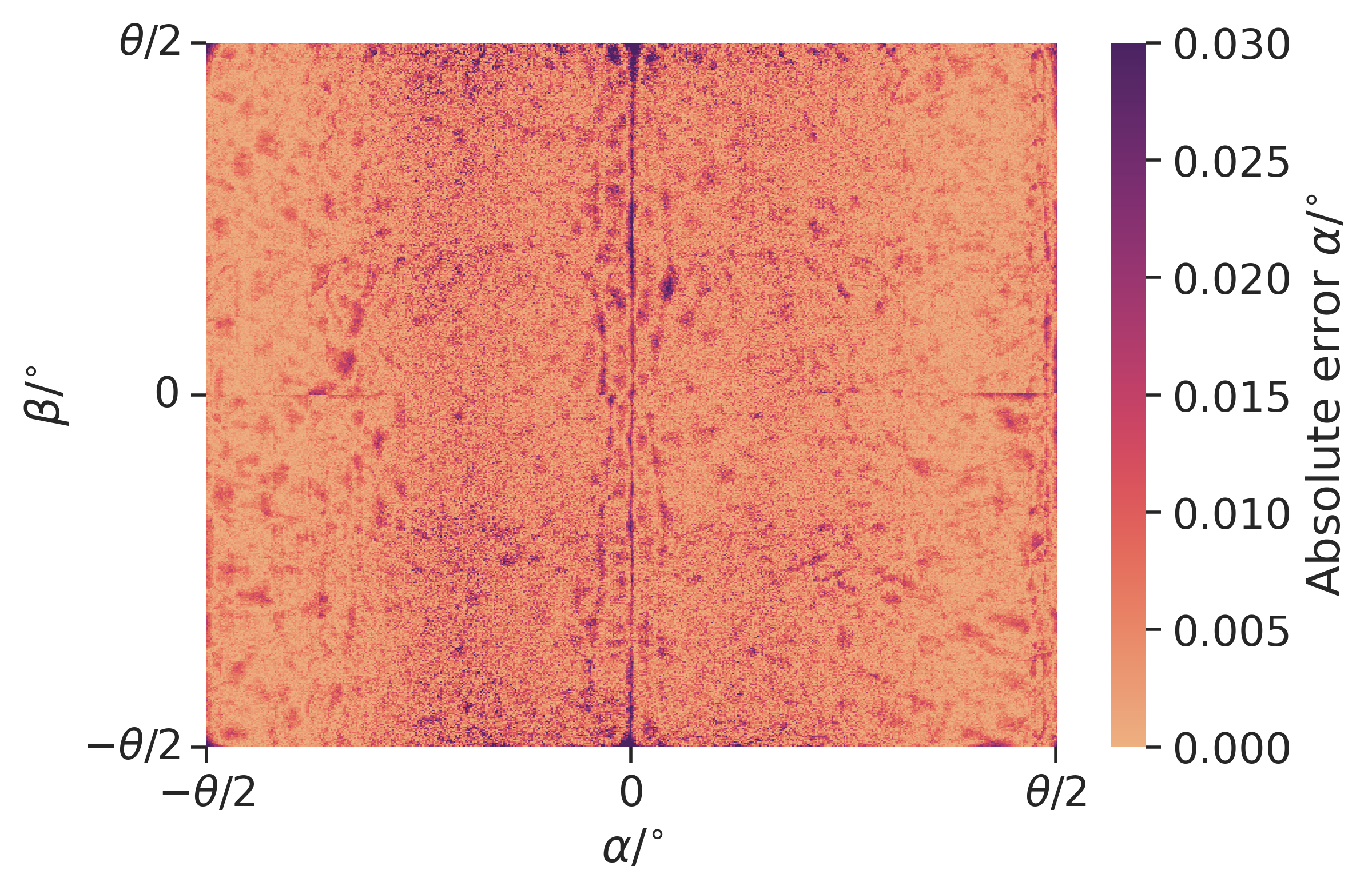}
        \caption{Absolute error heatmap for angle $\alpha$ in a multi-aperture configuration with non-augmented model using the entire dataset.}
        \label{fig:ma_hma}
    \end{subfigure}
    
    \begin{subfigure}[b]{0.95\textwidth}
        \centering
        \includegraphics[width=\textwidth]{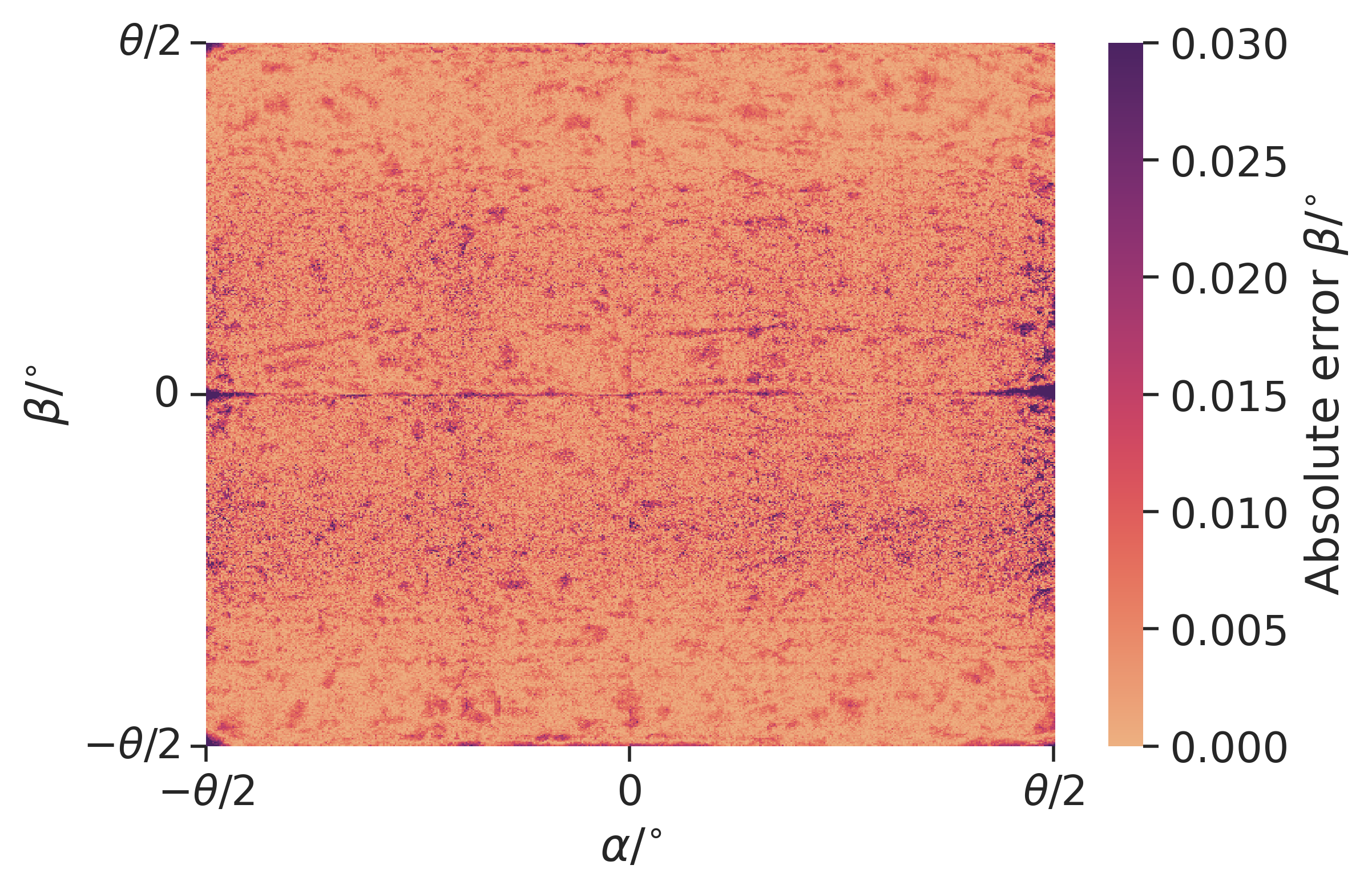}
        \caption{Absolute error heatmap for angle $\beta$ in a multi-aperture configuration with non-augmented model using the entire dataset.}
        \label{fig:ma_hmb}
    \end{subfigure}
    
    \caption{Absolute error heatmaps for multi-aperture configuration with non-augmented model.}
    \label{fig:ma_hm_overall}
\end{figure*}

\textbf{Absolute error heatmaps.} The performance of the multi-aperture non-augemented model is analyzed by investigating the two-axis absolute error landscape over the full sensor FOV as illustrated by the heatmaps in Figure \ref{fig:ma_hm_overall}. The absolute error heatmap for angle $\alpha$ is shown in Figure \ref{fig:ma_hma}, while the heatmap for angle $\beta$ is shown in Figure \ref{fig:ma_hmb}. The overall error over the two-axis FOV is low for both angles and appears to be smoother than the single-aperture heatmaps. Similar to the single-aperture case, the error tends to be around 0.005° as well. While there are still some small pockets of higher errors, they appear to be less numerous and more localized. Again, it is likely that training on a larger portion of the dataset would resolve these pockets.

Moreover, the errors at the corners of the FOV are largely resolved by the multi-aperture configuration as hypothesized. This makes sense, as there are more features to learn from even after the central aperture of the array is partially visible. In addition, while the class boundary error is still visible in both angle heatmaps, the effect appear to be greatly reduced for the multi-aperture case. Here, we note that the class uncertainty appears to instead follow a line of error along the zero line of the respective angular estimate. Nevertheless, we believe that this effect can be further mitigated by using even more apertures or relaxing the class weights with dropout to better generalize to unseen data.

\begin{figure*}[H]
    \centering
    \begin{subfigure}[b]{0.49\textwidth}
        \centering
        \includegraphics[width=\linewidth]{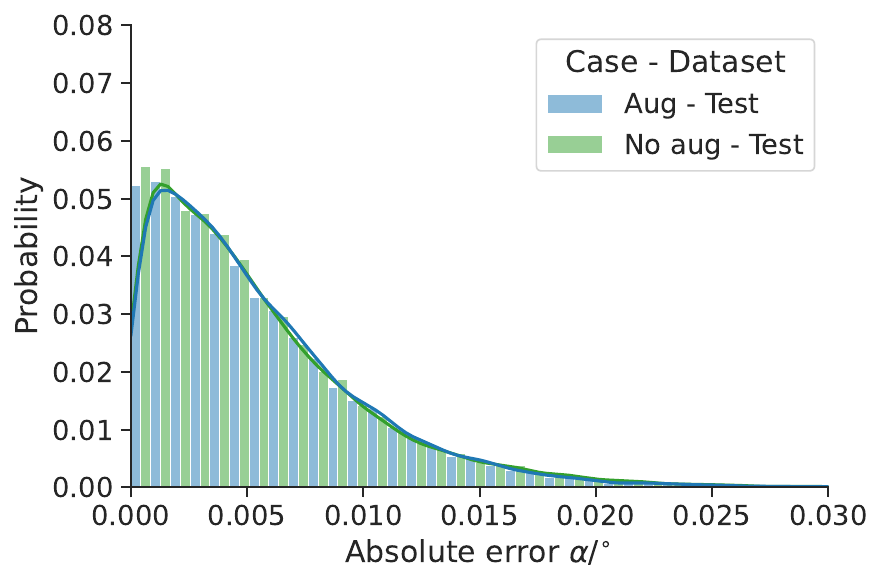}
        \caption{Absolute error histogram for angle $\alpha$ in a multi-aperture configuration with augmented and non-augmented model on the testing dataset.}
        \label{fig:ma_hista}
    \end{subfigure}
    \hfill
    \begin{subfigure}[b]{0.49\textwidth}
        \centering
        \includegraphics[width=\linewidth]{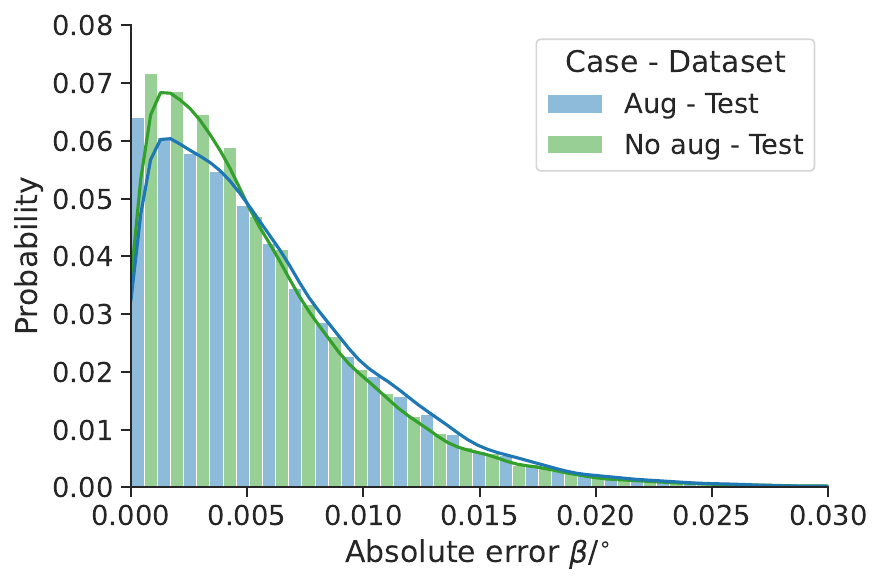}
        \caption{Absolute error histogram for angle $\beta$ in a multi-aperture configuration with augmented and non-augmented model on the testing dataset.}
        \label{fig:ma_histb}
    \end{subfigure}
    
    \vspace{0.25cm} 
    
    \begin{subfigure}[b]{0.49\textwidth}
        \centering
        \includegraphics[width=\linewidth]{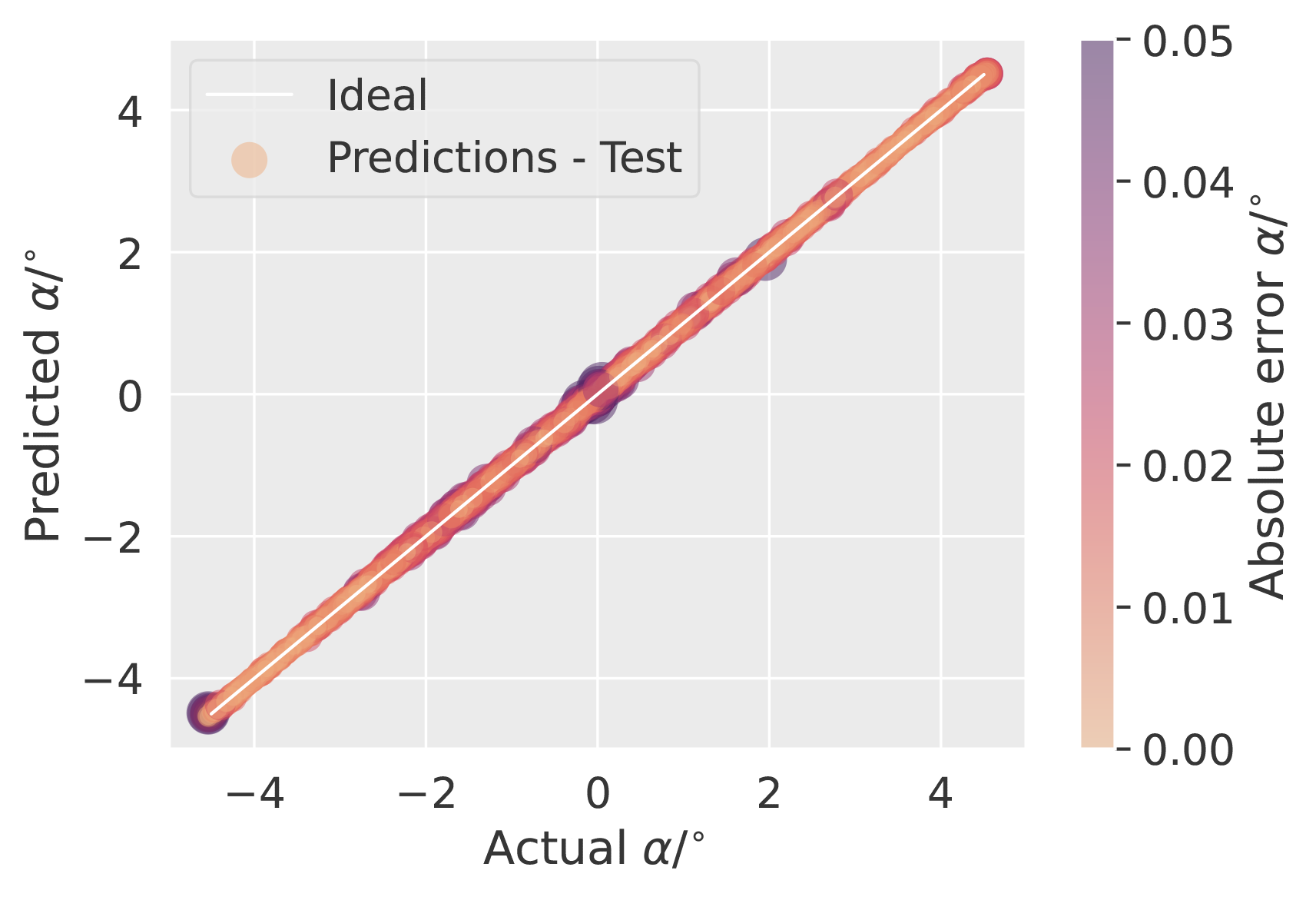}
        \caption{Predicted vs actual diagram of the absolute error for angle $\alpha$ in a multi-aperture configuration with non-augmented model on the testing dataset.}
        \label{fig:ma_pvaa}
    \end{subfigure}
    \hfill
    \begin{subfigure}[b]{0.49\textwidth}
        \centering
        \includegraphics[width=\linewidth]{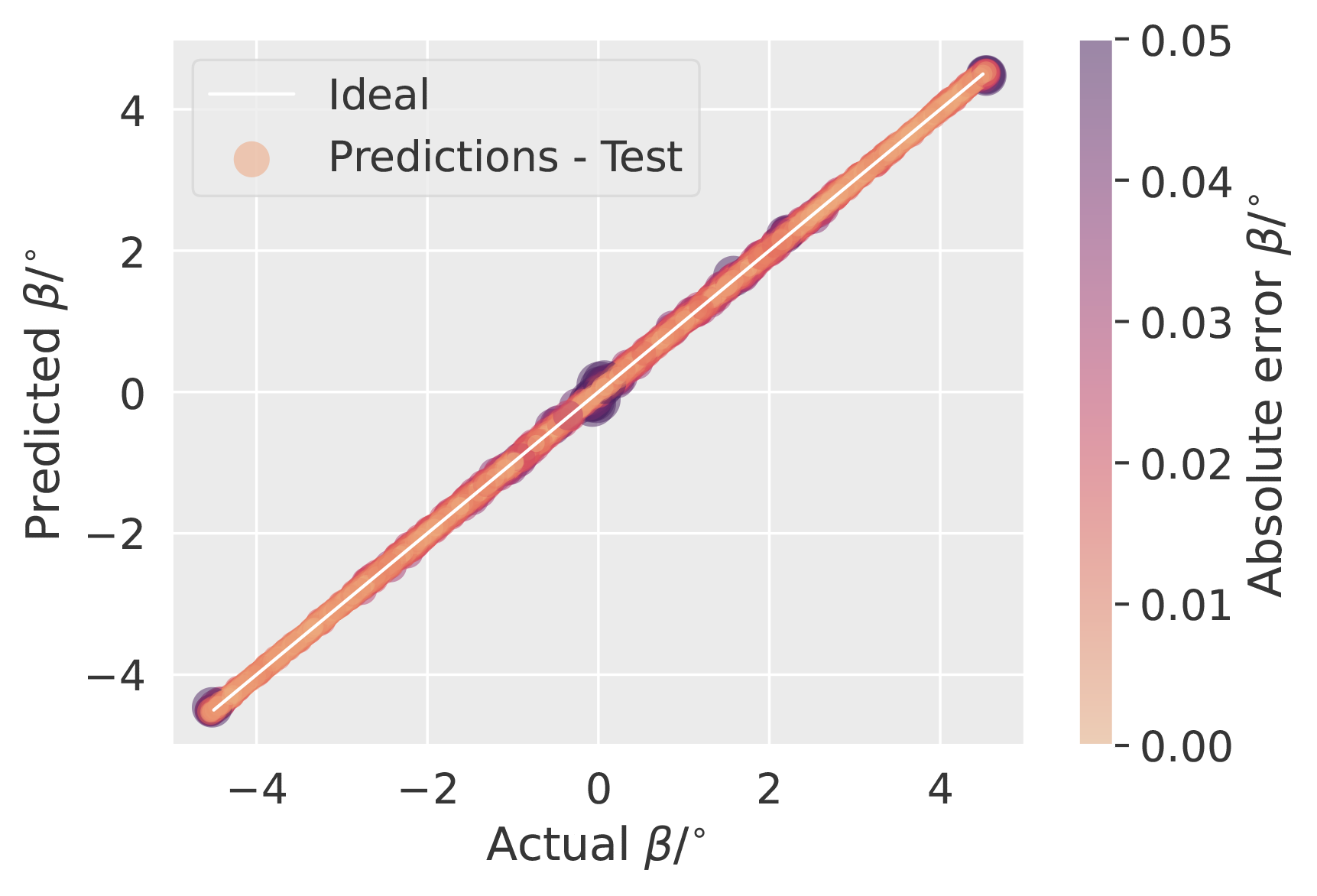}
        \caption{Predicted vs actual diagram of the absolute error for angle $\beta$ in a multi-aperture configuration with non-augmented model on the testing dataset}
        \label{fig:ma_pvab}
    \end{subfigure}
    
    \caption{Absolute error metrics of the multi-aperture configuration.}
    \label{fig:ma_error_overall}
\end{figure*}

\textbf{Absolute error histograms.} We now evaluate the absolute error metrics of the multi-aperture configuration in Figure \ref{fig:ma_error_overall}. To begin, the absolute error distributions via error histograms of the augmented and non-augmented models on the testing dataset are explored. The absolute error histogram for angle $\alpha$ is shown in Figure \ref{fig:ma_hista}, while the error histogram for angle $\beta$ is shown in Figure \ref{fig:ma_histb}. The histograms are displayed in a "dodge" format, in which the augmented and non-augmented distributions are plotted side-by-side. For consistency the augmented model is blue and the non-augmented model is green. Moreover, the histograms are displayed as probability distribution functions over the error bins.

Both of the incident angle estimates share similar distribution shapes, with the absolute error spread within 0.02° and mean of about 0.005°. As with the single-aperture case, the augmented and non-augmented models report similar performance on their respective datasets. However, the non-augmented model accuracy is slightly better due to the augmented model needing to filter out the learned noise and thresholding image corruption. The multi-aperture augmented model remains robust to the learned uncertainty with a small error discrepancy.

\textbf{Absolute error predicted vs actual.} The multi-aperture non-augmented model test predictions are evaluated using predicted vs actual diagrams in Figure \ref{fig:ma_error_overall}. The predicted vs actual diagram of the absolute error for angle $\alpha$ is presented in Figure \ref{fig:ma_pvaa}, while the predicted vs actual diagram of the absolute error for angle $\beta$ is presented in Figure \ref{fig:ma_pvab}. The white line portrays the ideal prediction trend, while the model test predictions are plotted as a scatter plot. The size and colormap of the scatter points vary with the magnitude of the absolute error. We find a good fit of the testing data to the actual targets. Unlike the single-aperture case, the multi-aperture errors tend to reside at the origin rather than the FOV corners. This error follows the heatmap trends presented earlier, where the FOV corner error was mitigated as compared to the single-aperture heatmap.

\subsection{Robustness evaluation}

\begin{figure*}[t]
    \centering
    \begin{subfigure}[b]{0.49\textwidth}
        \centering
        \includegraphics[width=\linewidth]{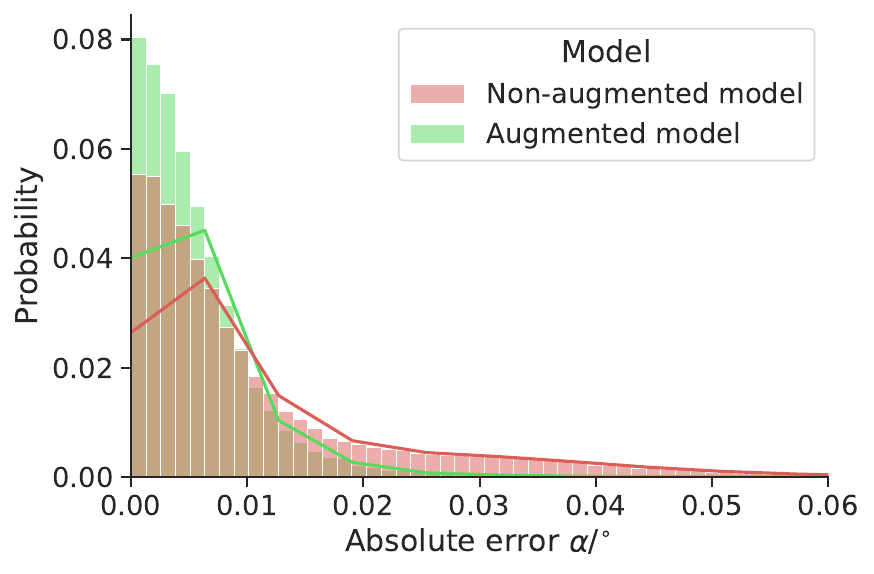}
        \caption{Comparison of non-augmented and augmented model performance on augmented testing data as an absolute error histogram for angle $\alpha$ in a single-aperture configuration.}
        \label{fig:sa_hista_noise}
    \end{subfigure}
    \hfill
    \begin{subfigure}[b]{0.49\textwidth}
        \centering
        \includegraphics[width=\linewidth]{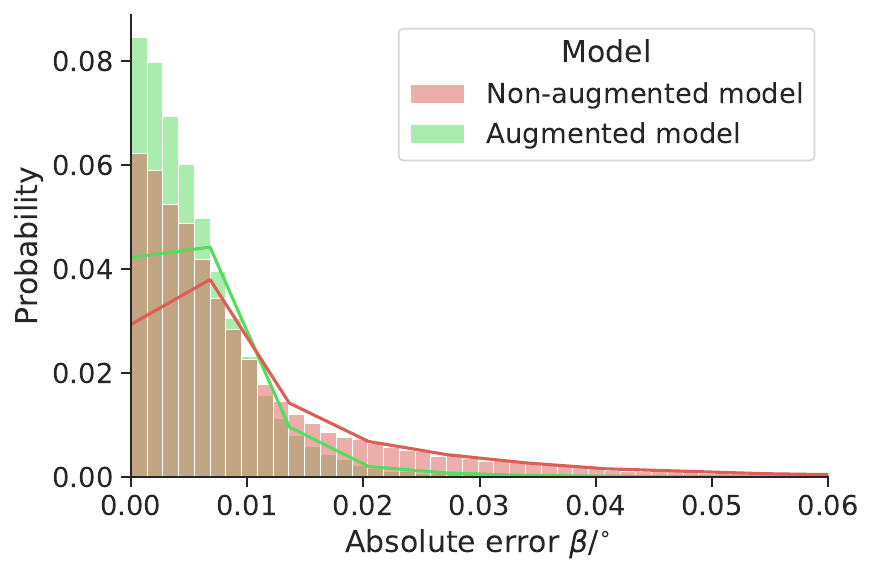}
        \caption{Comparison of non-augmented and augmented model performance on augmented testing data as an absolute error histogram for angle $\beta$ in a single-aperture configuration.}
        \label{fig:sa_histb_noise}
    \end{subfigure}
    
    \vspace{0.25cm} 
    
    \begin{subfigure}[b]{0.49\textwidth}
        \centering
        \includegraphics[width=\linewidth]{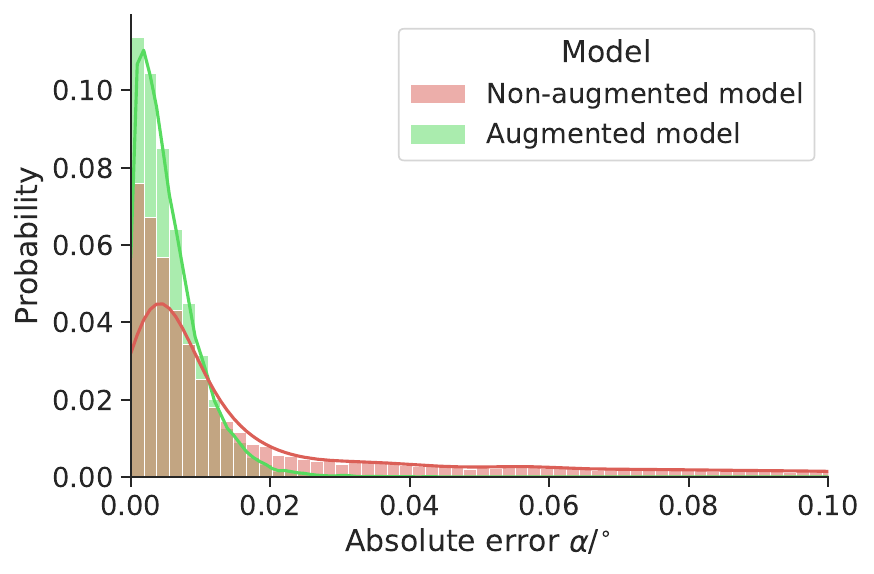}
        \caption{Comparison of non-augmented and augmented model performance on augmented testing data as an absolute error histogram for angle $\alpha$ in a multi-aperture configuration.}
        \label{fig:ma_hista_noise}
    \end{subfigure}
    \hfill
    \begin{subfigure}[b]{0.49\textwidth}
        \centering
        \includegraphics[width=\linewidth]{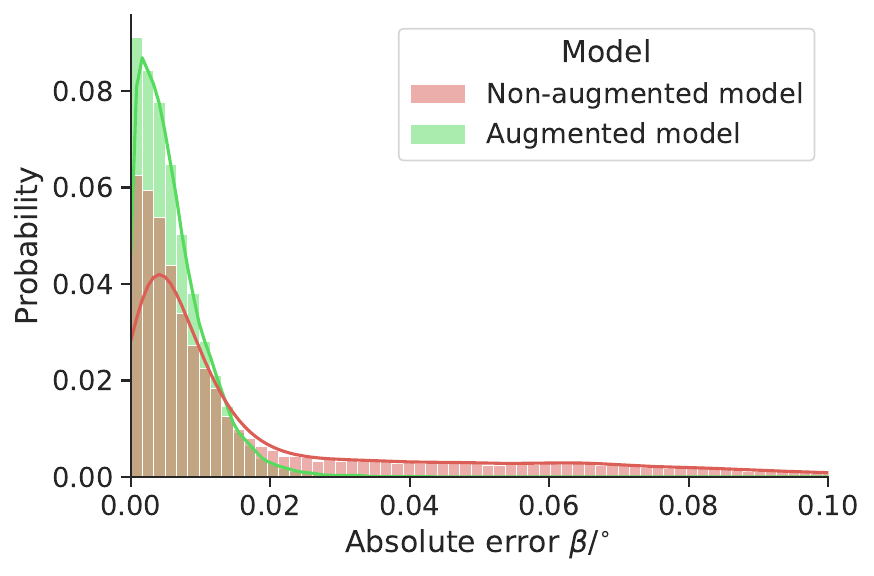}
        \caption{Comparison of non-augmented and augmented model performance on augmented testing data as an absolute error histogram for angle $\beta$ in a multi-aperture configuration.}
        \label{fig:ma_histb_noise}
    \end{subfigure}
    
    \caption{Comparison of non-augmented and augmented model performance on augmented testing data.}
    \label{fig:hist_noise}
\end{figure*}

In order to demonstrate that the proposed modeling approach is valid for operations in the harsh space environment, we must fully evaluate model performance on realistic image imperfections and corruption via the augmented data. To this end, we compare the performance of the augmented and non-augmented models on the augmented (noise + thresholding) dataset. By comparing the two models we can better understand just how well the augmented model learned the noise and thresholding features as compared to the non-augmented baseline. For this study, we compare the model robustness via layered absolute error histograms as presented in Figure \ref{fig:hist_noise}.

However, before discussing the performance comparison, we first must detail out the challenges typically associated with the augmentation added. The image thresholding used is an upper cutoff of intensity to emulate pixel saturation typically encountered during operations. In addition, the noise added to the augmented images is a combination of Poisson distributed shot-noise and Gaussian distributed read-noise as outlined in Table \ref{tab:detectorparam}. Unlike Gaussian noise which is additive, Poisson noise is signal dependent and therefore more challenging to filter. Luckily, CNNs are highly capable of implicitly learning complex noise patterns and handling mixed noise types, such as Poisson-Gaussian noise. This attribute greatly improves the model robustness and accuracy to real-world applications.

Next, we directly compare the model robustness of the non-augmented and augmented models to the augmentation dataset via layered absolute error histograms. The single-aperture robustness is evaluated for angle $\alpha$ in Figure \ref{fig:sa_hista_noise} and for angle $\beta$ in Figure \ref{fig:sa_histb_noise}. For the single-aperture configuration we see a clear trend of the augmented model outperforming the non-augmented model for both $\alpha$ and $\beta$. The augmented model error distribution has a mean error of 0.006° and a spread of 0.03°, whereas the non-augmented model has a mean error of 0.01° and a spread of 0.06°. Hence, the augmented model has demonstrated the ability to learn the corrupted image features and remain robust under uncertainty.

Furthermore, the multi-aperture robustness is evaluated for angle $\alpha$ in Figure \ref{fig:ma_hista_noise} and for angle $\beta$ in Figure \ref{fig:ma_histb_noise}. For the multi-aperture configuration we see the same trend of the augmented model beating the non-augmented model for both $\alpha$ and $\beta$. Here, the augmented model error distribution has a mean error of 0.006° and a spread of 0.03°, whereas the non-augmented model has a mean error of 0.02° and a spread of 0.1°. Surprisingly, we again conclude that the single-aperture augmented model appears to be as robust as the multi-aperture augmented model to image corruption, thereby nullifying the conventional advantages of multi-aperture configurations. Similar to the single-aperture case, the multi-aperture augmented model has also demonstrated the ability to learn the corrupted image features and remain robust under uncertainty.

\subsection{Results summary}

This concludes the presented results of the single-aperture and multi-aperture SSCNN networks under augmented and non-augmented data. We summarize the section by highlighting key findings from the results:

\begin{enumerate}[nolistsep]
    \item \textbf{Single-aperture error landscape:} The highest errors are at the corners of the FOV and at the sub-FOV class boundaries. The class boundary uncertainty is derived from the perception of the single aperture as within two or more classes simultaneously. 
    \item \textbf{Single-aperture class boundaries:} The class uncertainty follows the sub-FOV quadrant boundaries.
    \item \textbf{Multi-aperture error landscape:} The errors at the corners of the FOV are reduced compared with single-aperture since there are more features to learn from even after the central aperture of the array is partially visible. In addition, the class boundary errors are also reduced since the apertures are less likely to be perceived as within two or more classes at the same time with multiple apertures.
    \item \textbf{Multi-aperture class boundaries:} The class uncertainty appears to follow a line of error along the zero line of the respective angular estimate.
    \item \textbf{Error metrics:} The error spread is within 0.02° and mean of is 0.005°. The augmented and non-augmented models display similar performance on their respective datasets.
    \item \textbf{Robustness:} Our proposed model is highly capable of implicitly learning complex noise patterns and handling mixed noise types, thereby greatly improving the model robustness and accuracy to real-world applications.
\end{enumerate}

\section{Discussion}\label{sec:discuss}

In this section, we discuss key findings, limitations, and future research directions.

\subsection{Limitations}

While the proposed methodology has numerous advantages compared with traditional digital sun sensor calibration techniques, it is not without potential limitations. These identified limitations include:

\begin{enumerate}[nolistsep]
\item \textbf{Validation challenges.} It is difficult to validate the proposed methodology against traditional calibration techniques since there are no baseline end-to-end fused models to compare against. Feature extraction or model representation methods alone are not directly comparable since our proposed SSCNN method neither calculates centroid features nor formulates the sensor response explicitly. Therefore, our validation approach for this study is primarily demonstrated through rigorous testing and robustness analyses.

\item \textbf{Data availability.} Training CNN models requires massive datasets for adequate convergence. While we offer a physics-informed data generation framework to remedy this, the challenges of limited data are evident. This is especially true since each unique sensor may require its own dataset for optimal performance.

\item \textbf{Computational complexity.} Low processing latency is critical for on-board AI estimation techniques. While SSCNNs offer greatly improved interference speed as compared with their dense counterparts, we did not specifically benchmark the model efficiency for real-time operations.

\item \textbf{Explainability.} Deep learning models are inherently opaque, thereby making trustworthy and safe estimates difficult to guarantee. Our proposed methodology is no exception. While much is gained from an end-to-end fused learning approach, the model weights are not physically interpretable parameters like conventional sun sensor models.

\item \textbf{Sim2Real gap.} This gap is defined as the difference between the true and simulated image data. To bridge the gap we must account for the primary sources of uncertainty in digital sun sensor modeling. For the digital sun sensor these sources are derived from the detector, mask, and light source. For this study, we emphasize that the simulated dataset is for pre-training and model prototyping purposes only, and it is not considered sufficient for in-flight use. One way of improving the Sim2Real gap is by applying transfer learning to the proposed pre-trained model with real sensor data. We expand on this idea in Section \ref{sec:futurework}.
\end{enumerate}

\subsection{Future Work} \label{sec:futurework}

The many challenges and successes throughout the development of this work have assisted in identifying new avenues for future research. We outline recommended future research directions for SSCNN-based digital sun sensor predictive calibration below:

\textbf{Superresolution.} The loss floor is most likely limited by the pixel quantization of the feature space rather than noise errors. Superresolution techniques could lead to improved accuracy if implemented.

\textbf{Scalability.} We intentionally limited the FOV of sun angles to be estimated in this study, as it was not required to demonstrate the transformative performance gains over conventional calibration approaches. Our network has proven capable of implicitly learning the effects of a larger boresight angle (diffraction, spot deformation, blurring, etc). However, we recommend scaling the implementation up to a larger FOV as next steps. Of course this comes with the challenges of more data and longer training times.

\textbf{Architecture optimization.} We settled on the ResNet-34 backbone for our architecture since it was found that a well-tuned general-purpose network can yield results close to state-of-the-art \cite{Lathuiliere2020}. Nevertheless, it is of interest to further optimize the architecture, especially for specific use-cases such as maximizing accuracy or minimizing complexity. We recommend experimenting with both deeper networks for improved accuracy and shallower networks for real-time edge operations.

\textbf{Edge computing.} Another major milestone for this research is to demonstrate real-time operations via onboard AI edge devices. While space edge computing and on-board AI are still relatively nascent, they are becoming closer to their terrestrial mobile edge computing counterparts (MEC) \cite{Kothari2020}. Improvements to space hardware and deep learning algorithms have proven sufficient for deployment in space for real-time operations.

In the work by Buonaiuto et al. \cite{Buonaiuto2017} a Nvidia-TX1 is used for satellite identification algorithms. The AMD-Xilinx Versal FPGA development board is used in the work by Petry et al. \cite{Petry2023} to run CNNs for real-time processing of images in the space domain. CloudScout by Giuffrida et al. \cite{Giuffrida2020} uses a Myriad 2 Vision Processing Unit (VPU) for real-time CNN processing for cloud classification in hyperspectral images. Finally, Kikuya et al. \cite{Kikuya2023} use a deep learning algorithm for real-time operations of an Earth camera, where the Raspberry Pi series of boards were evaluated. For future directions, we recommend the investigation of FPGAs like the Xilinx Versal for use with the proposed methodology since the accelerators of the Versal are optimized for CNNs.

\textbf{Bridging the Sim2Real gap.} The Sim2Real gap can be accounted for by implementing transfer learning. Transfer learning updates the weights of the pre-trained model with features specific to the real sensor response. By fine-tuning the proposed pre-trained model on a small set of real sensor data the final model performance and overall accuracy can be improved \cite{Allworth2021}. Furthermore, it was found by \citet{Allworth2021} that models pre-trained on simulated data and then fine-tuned on real-world data outperformed models that were only trained on real data alone. Therefore, we recommend using transfer learning to pre-train a model using the proposed framework on synthetic data and then fine-tune on real sensor data to achieve comparable performance to a full real sensor dataset, thereby overcoming the Sim2Real gap.

\section{Conclusion}\label{sec:conc}

In this study, we developed and tested a novel end-to-end predictive calibration framework for digital sun sensors using a sparse submanifold convolutional neural network. We introduce a publicly available physics-informed synthetic dataset that incorporates diffraction and real-sensor noise. Both single and multi-aperture mask configurations are selected as the illumination patterns for the study. By leveraging a synthetic dataset, the calibration methodology can be accelerated.

The network is based on the ResNet-34 backbone modified with sparse convolutional variants. It was found that the use of sparse convolutional layers was mandatory for convergence due to the sparsity of the data. Moreover, we implemented sparse submanifold convolutional layers (SSCNNs) so that only the non-sparse pixels were processed, thereby greatly improving the computational efficiency. The network was trained using the MSE loss function and Adamax optimizer. After training concluded, the final scaled loss converged to around $2\mathrm{e}-6$ for both the single and multi-aperture models.

Model credibility was assessed through training validation, testing, and robustness evaluation. We selected MSE loss and R-squared as the training validation metrics. The accuracy of the network is competitive with state of the art ultra-fine sun sensor approaches, while requiring less apertures for the same performance. The simulated results suggest an achievable mean error of around 0.005° for the notional sensor configuration. While evaluating for robustness, we found the network to demonstrate impressive learning ability of complex and layered uncertainty sources. However, the proposed method should be further tested on real sensor data to explore the real-world returns of the calibration framework.

Advantages of our proposed methodology include: data-driven decision making to formulate the model, separate feature extraction and model representation is not required, no segmentation is required due to sparse layers, the approach is fully mask agnostic, richer feature extraction is enabled through convolutional layers, no pre-process denoising is required, and correlation for sub-FOVs is supported via classes. The main contributions of this work are: (1) the first time to our knowledge that a CNN regression model is applied to the problem of sun sensor predictive calibration, (2) introducing an end-to-end training approach for a single SSCNN, designed to concurrently handle feature extraction and correlation mapping, (3) generating a synthetic dataset of physics-informed digital sun sensor images, augmented with real sensor noise and thresholding, for training purposes, and (4) conducting a performance evaluation of the deep learning approach to various mask configurations. Through this work, we hope to empower future researchers and practitioners with a new framework to more accurately and efficiently calibrate digital sun sensors of all mask types.

\appendix
\section{Data Availability} \label{sec:appendixa}

The following data and software resources used in this study are publicly available on Zenodo:
\begin{itemize}
  \item The \textbf{image dataset} compiled for training and evaluation is available at \href{https://doi.org/10.5281/zenodo.15778886}{Image Dataset} \cite{Herman2025d}.
  \item The \textbf{sun sensor simulation tool} is available at \href{https://doi.org/10.5281/zenodo.16579563}{Sun Sensor Simulation} \cite{Herman2025g}.
  \item The \textbf{model training and testing code} is available at \href{https://doi.org/10.5281/zenodo.16579535}{Model Code} \cite{Herman2025f}.
  \item The \textbf{trained calibration models} are available at \href{https://doi.org/10.5281/zenodo.15778990}{Trained Models} \cite{Herman2025e}.
\end{itemize}


\bibliographystyle{model1-num-names}

\bibliography{cas-refs}





\end{document}